\documentclass[twocolumn,notitlepage,prl,superscriptaddress,citeautoscript,showpacs,amsmath,amstex,amssymb,mathfonts,showpacks]{revtex4-1}

\pdfoutput=1

\usepackage{amsmath}
\usepackage{amssymb}
\usepackage{graphicx}
\usepackage{appendix}
\usepackage{enumerate}
\usepackage{setspace}
\usepackage{mathrsfs}
\usepackage{color}

\usepackage{subfigure}

\usepackage[protrusion=true,expansion=true]{microtype}

\usepackage{hyperref}
\hypersetup{
	pdftoolbar=true,        		 % show AcrobatÕs toolbar?
	pdfmenubar=true,     		 % show AcrobatÕs menu?
	pdffitwindow=false,     	 % window fit to page when opened
	pdfstartview={FitH},    	 % fits the width of the page to the window
	pdftitle={Scaling Theory of Entanglement at the Many-Body Localization Transition},   % title
	pdfauthor={Dumitrescu, Vasseur, Potter},     		 % author
	pdfsubject={},   			 % subject of the document
	pdfcreator={},   	        	   	 % creator of the document
	pdfproducer={}, 		% producer of the document
	pdfnewwindow=true,      	% links in new window
	colorlinks=true,       		% false: boxed links; true: colored links
	linkcolor=black,          	% color of internal links (change box color with linkbordercolor)
	citecolor=blue,        		% color of links to bibliography
	filecolor=magenta,    		% color of file links
	urlcolor=blue          		% color of external links
}

\DeclareMathOperator{\sgn}{sgn}

\newcommand{\e}{\varepsilon}

\renewcommand{\(}{\left(}
\renewcommand{\)}{\right)}
\renewcommand{\[}{\left[}
\renewcommand{\]}{\right]}

\renewcommand{\d}{\partial}

\newcommand{\header}[1]{\noindent{\bf#1 -- }}

\begin{document}

\author{Philipp T.~Dumitrescu}
\email{philippd@utexas.edu}
\affiliation{Department of Physics, University of Texas at Austin, Austin, Texas 78712, USA}

\author{Romain Vasseur}
\affiliation{Department of Physics, University of California, Berkeley, California 94720, USA}
\affiliation{Materials Science Division, Lawrence Berkeley National Laboratories, Berkeley, California 94720, USA}
\affiliation{Department of Physics, University of Massachusetts, Amherst, Massachusetts 01003, USA}

\author{Andrew C.~Potter}
\affiliation{Department of Physics, University of Texas at Austin, Austin, Texas 78712, USA}

\title{Scaling Theory of Entanglement at the Many-Body Localization Transition}
\date{\today}

\begin{abstract}
We study the universal properties of eigenstate entanglement entropy across the transition between many-body localized (MBL) and thermal phases. We develop an improved real space renormalization group approach that enables numerical simulation of large system  sizes and systematic extrapolation to the infinite system size limit. For systems smaller than the correlation length, the average entanglement follows a sub-thermal volume law, whose coefficient is a universal scaling function. 
The full distribution of entanglement follows a universal scaling form, and exhibits a bimodal structure that produces universal subleading power-law corrections to the leading volume-law.  For systems larger than the correlation length, the short interval entanglement exhibits a discontinuous jump at the transition from fully thermal volume-law on the thermal side, to pure area-law on the MBL side.
\end{abstract}


\maketitle

Recent experimental advances in synthesizing isolated quantum many-body systems, such as cold-atoms~\cite{bloch_review,Schreiber842,choi2016exploring,2016arXiv161207173L}, trapped ions~\cite{Smith2016aa,2016arXiv160908684Z}, or impurity spins in solids~\cite{2016arXiv161008057C,2016arXiv161205249W}, have raised fundamental questions about the nature of statistical mechanics. Even when decoupled from external sources of dissipation, large interacting quantum systems tend to act as their own heat-baths and reach thermal equilibrium. This behavior is formalized in the eigenstate thermalization hypothesis (ETH)~\cite{PhysRevE.50.888,PhysRevA.43.2046}. Generic excited eigenstates of such thermal systems are highly entangled, with the entanglement of a subregion scaling as the volume of that region (``volume law").  This results in incoherent, classical dynamics at long times.  In contrast, strong disorder can dramatically alter this picture by pinning excitations that would otherwise propagate heat and entanglement~\cite{PhysRev.109.1492,FleishmanAnderson,Gornyi,BAA,PhysRevB.75.155111,PalHuse,BauerNayak}. In such many-body localized (MBL) systems~\cite{2014arXiv1404.0686N,Altman:2015aa,1742-5468-2016-6-064010}, generic eigenstates have properties akin to those of ground states. They exhibit short-range entanglement that scales like the perimeter of the subregion~\cite{BauerNayak} (``area law"), and have quantum coherent dynamics up to arbitrarily long time scales~\cite{PhysRevB.77.064426,PhysRevLett.109.017202,PhysRevLett.110.260601,PhysRevLett.113.147204,PhysRevB.91.140202,BahriMBLSPT,PhysRevB.90.174302}, even at high energy densities~\cite{HuseMBLQuantumOrder, BauerNayak,BahriMBLSPT,PhysRevB.89.144201,PekkerRSRGX,PhysRevLett.113.107204}.

A transition between MBL and thermal regimes requires a singular rearrangement of eigenstates from area-law to volume-law entanglement. This many-body (de)localization transition (MBLT) represents an entirely new class of critical phenomena, outside the conventional framework of equilibrium thermal or quantum phase transitions. Developing a systematic theory of this transition promises not only to expand our understanding of possible critical phenomena, but also to yield universal insights into the nature of the proximate MBL and thermal phases.

The eigenstate entanglement entropy can be viewed as a non-equilibrium analog of the thermodynamic free energy for a conventional thermal phase transition, and plays a central role in our conceptual understanding of the MBL and ETH phases. Describing the entanglement across the MBLT requires addressing the challenging combination of disorder, interactions, and dynamics. Consequently, most studies have resorted to fully microscopic simulation methods like exact diagonalization (ED)\cite{PhysRevB.75.155111,PalHuse,Luitz,PhysRevB.93.060201,khemani2016critical}. The exponential complexity of such methods fundamentally limits them to small systems ($\lesssim 30$ sites), preventing them from accurately capturing universal scaling properties. For example, critical exponents computed from ED violate rigorous scaling bounds~\cite{Chayes,chandran2015finite}. 

A promising alternative is to eschew a microscopic description, which is not required to compute universal scaling properties, and instead develop a coarse grained renormalization group (RG) description. Two related RG approaches~\cite{Vosk:2015aa,Potter:2015aa} have produced a consistent picture of the MBLT (see also~\cite{PhysRevB.93.224201}). Nonetheless, both approaches rest on ad-hoc albeit plausible heuristics for computing many-body matrix elements. In this paper, we develop a RG scheme building upon  \cite{Potter:2015aa}, but whose steps are rooted in well-established properties of matrix elements in MBL and thermal systems. Using this modified RG scheme, we compute the full scaling structure of entanglement across the transition, by simulating large systems sizes with many ($10^{5}$--$10^{6}$) disorder realizations, that allow systematic extrapolation to the infinite size limit. The resulting scaling properties depart dramatically from those of conventional equilibrium critical points, highlighting the unusual nature of the MBLT.

\header{RG approach}
Our RG approach builds a coarse-grained picture of eigenstates by identifying collective many-body resonances that destabilize the MBL phase. Although this approach is not tied to a particular microscopic model, we picture a chain of spinless fermions with Hamiltonian 
$
H=\sum_x( -c^{\dagger}_xc^{\vphantom\dagger}_{x+1}+\text{H.c.}-\mu_x \rho_x+V\rho_x\rho_{x+1}),
$
Here $\rho_{x} =  c^{\dagger}_{x} c^{\vphantom\dagger}_{x}$ is the fermion density on site $x$, and $\mu_x$ is a random chemical potential drawn from a uniform distribution on $[0,W]$. 
The noninteracting system ($V=0$) is Anderson localized with localization length $x_0 \approx 2 / \log\(1 + W ^2 \)$~\cite{Potter:2015aa}. Interactions ($|V|>0$) can drive multiparticle collective resonances. For weak interactions, $V\ll W$, the system remains MBL and these resonances restructure the local integrals of motion (LIOMs) from weakly dressed single-particle orbitals to few-body LIOMs~\cite{PhysRevLett.111.127201,PhysRevB.90.174202,Ros2015420,PhysRevLett.117.027201}.  For sufficiently strong interactions, MBL breaks down as all degrees of freedom resonate.

While finding the true resonances is tantamount to solving the many-body Hamiltonian, close to the continuous MBLT, one expects a scale-invariant structure in which resonances are organized hierarchically and can be constructed iteratively~\cite{Potter:2015aa}.
Since large many-fermion resonances will drive the MBLT, it is natural to consider an effective model in terms of resonant clusters,
i.e.~groups of inter-resonating single-particle orbitals, characterized only by coarse grained information: the effective bandwidth $\Lambda_i$ and the typical level spacing $\delta_i$.

To characterize cluster interactions, we retain only the typical amplitude $\Gamma_{ij}$ of matrix elements for
transitions changing the states of clusters $i$ and $j$, and compare this to the corresponding typical energy mismatch $\Delta E_{ij}$ between those states. 
For $\Gamma_{ij} \gg \Delta E_{ij}$, states of $i$ and $j$ will resonantly admix, whereas for $\Gamma_{ij}\ll \Delta E_{ij}$, the clusters will remain decoupled apart from weak virtual dressing. 
We divide these regimes sharply and define a resonant coupling if $\Gamma_{ij}  > \Delta E_{ij}$. The ambiguity of this partition 
becomes unimportant for the large clusters determining the transition, since both $\Gamma_{ij}$ and $\Delta E_{ij}$ depend exponentially on fluctuating extensive quantities, and are rarely comparable.

The RG procedure for a chain of $L$ sites with periodic boundary conditions proceeds as follows.
Initially, each cluster corresponds to a localized single-particle orbital with bandwidth $\Lambda_i = \e_i\approx \mu_i$ ($\e_i$ the non-interacting single-particle energy), $\Delta E_{ij} = \vert \mu_i - \mu_j \vert$, and $\Gamma_{ij} = V (e^{- \vert i - j \vert / x_0} + e^{- \vert i - j  - L\vert / x_0} )$. {We set $V=0.3$ throughout.}
During an RG step, all clusters connected by a path of resonating bonds are merged into a new cluster $\{i\} \to i'$. 
The coarse grained parameters of the newly formed cluster are chosen as~\footnote{Unless the level spacing of one cluster exceeds the bandwidth of the other; then $\Lambda_{i'} \geq \delta_{i'} \geq \Lambda_{j'} \geq \delta_{j'}$ and $\Delta E_{i'j'} = \max\left(\delta_{i'} - \Lambda_{j'}, \delta_{j'}\right)$}:  
$\Lambda_{i'} = [{{\sum_{i} \Lambda^2_{i} + \sum_{ij} \Gamma^2_{ij}}}]^{1/2}$,
$\delta_{i'} = \Lambda_{i'} / (2^{n_{i'}} -1)$, and
$\Delta E_{i'j'} = \delta_{i'} \delta_{j'} / \min(\Lambda_{i'}, \Lambda_{j'})$
where $n_{i'}$ is the number of sites in cluster $i'$. 

\begin{figure}[t]
\centering
\includegraphics[width=0.6\columnwidth]{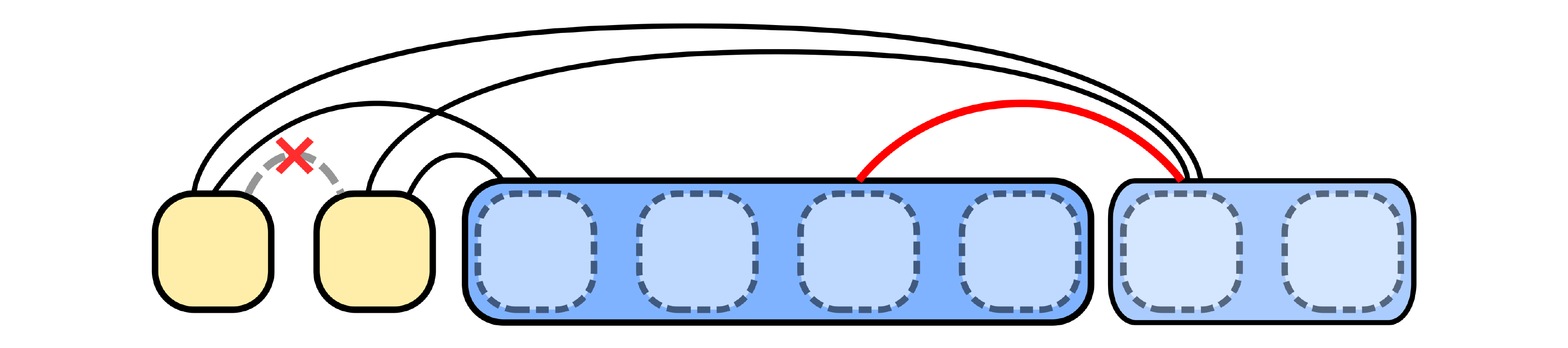}
\caption{{{\bf Schematic of a RG step. -- } \label{fig:RGrules} 
Eight initial clusters (dashed squares) interact with each other; those connected by a resonant path ($\Gamma_{ij} > \Delta E_{ij}$) merge into bigger, new clusters (colored rectangles). 
The coupling between new clusters is turned off or renormalized from the previous step (see text).
}}
\vspace{-0.2in}
\end{figure}

The effective inter-cluster couplings are changed according to two distinct rules, locally mirroring MBL or ETH behavior (Fig.~\ref{fig:RGrules}). 
First, consider two clusters not modified during a RG step. In isolation, these clusters would form a small MBL region, with decoupled LIOMs that project onto the separate states of each cluster. Any further resonance between these two clusters must be mediated by other clusters; we can therefore neglect the direct coupling between them and set $\Gamma_{i'j'} = 0$.
Second, if at least one of the clusters is modified during the RG step, the new coupling between two clusters is \cite{SupMat}
\begin{align}\label{eq:matRGrule}
\Gamma_{i'j'} = \[\max_{i_1\in \{i\}, i_2 \in \{j\}}\Gamma_{ij}\]e^{-(n_{i'}+n_{j'}-n_{i_1}-n_{i_2})s_\text{th}/2}.
\end{align}
Here, $\max \Gamma $ selects the strongest resonating pathway. The exponential factor approximates the resonating clusters as small locally thermal sub-systems with entropy $s_\text{th} = \log 2$ per site. This form holds for matrix elements of local operators in a finite-size, ETH system \cite{PhysRevE.50.888, SupMat}.

The renormalization of intercluster couplings is different from those of \cite{Vosk:2015aa,Potter:2015aa}, but have similarities to those of~\cite{2016arXiv160801815D}.
The coupling $\Gamma_{ij}$ sets the timescale over which clusters can resonate to change each other's state. 
Early in the RG, resonances are fast and occur directly between a few strongly coupled sites. Later in the RG, resonances are more collective and involve many sites.
Although the direct coupling $\Gamma_{ij}$ is set to zero if two clusters cannot resonate at a given time scale, they can still resonate later, if mediated via coupling to other clusters \cite{SupMat}.

Approximating $\Gamma_{ij}$ by the limiting MBL and ETH forms becomes self-consistently justified since the width of the distribution of resonance parameters $g_{ij} = \Gamma_{ij} / \Delta E_{ij}$~\cite{Vosk:2015aa,Potter:2015aa,PhysRevX.5.041047} increases with each RG step. In an infinite critical system, the width of the distribution of $g$ increases without bound along the RG flow so that one asymptotically encounters only the cases $g\ll 1$ (MBL) or $g \gg 1$ (ETH) and almost never faces marginal cases where $g\approx 1$. This flow to infinite randomness of $g$ justifies the RG approximations in an analogous fashion to other microscopic RG approaches for quantum phase transitions in disordered spin chains~\cite{FisherRSRG1,FisherRSRG2,PhysRevLett.112.217204,PekkerRSRGX,QCGPRL}. 

Rooting the $\Gamma_{ij}$ renormalization in well-established asymptotic properties more accurately captures the competition between locally MBL regions being thermalized by nearby locally thermal clusters, or isolating them.
 These rules cleanly prevent unphysical ``avalanche" instabilities of the MBL phase~\cite{2016arXiv160801815D,2016arXiv161100770A} in which an atypically large resonant cluster becomes increasingly thermal as it grows, enabling it to thermalize an arbitrarily large MBL region~\footnote{Re-examining~\cite{Potter:2015aa} showed that those RG rules allowed rare avalanches; this does not quantitively alter the reported critical properties.}.

The RG terminates if no resonant bonds remain or the system fully thermalizes. 
Like~\cite{Potter:2015aa}, our approach allows for a distribution of various cluster sizes in the final configuration. This feature is important, as typical configurations at criticality are predominantly MBL with few large clusters~\cite{Potter:2015aa} -- a picture supported by recent ED numerics~\cite{khemani2016critical}.   In contrast, the approach of \cite{Vosk:2015aa} allowed both MBL and thermal blocks (clusters) to grow until the system is one large block that is either thermal or MBL.

\begin{figure}[t]
\centering
\includegraphics[width=\columnwidth]{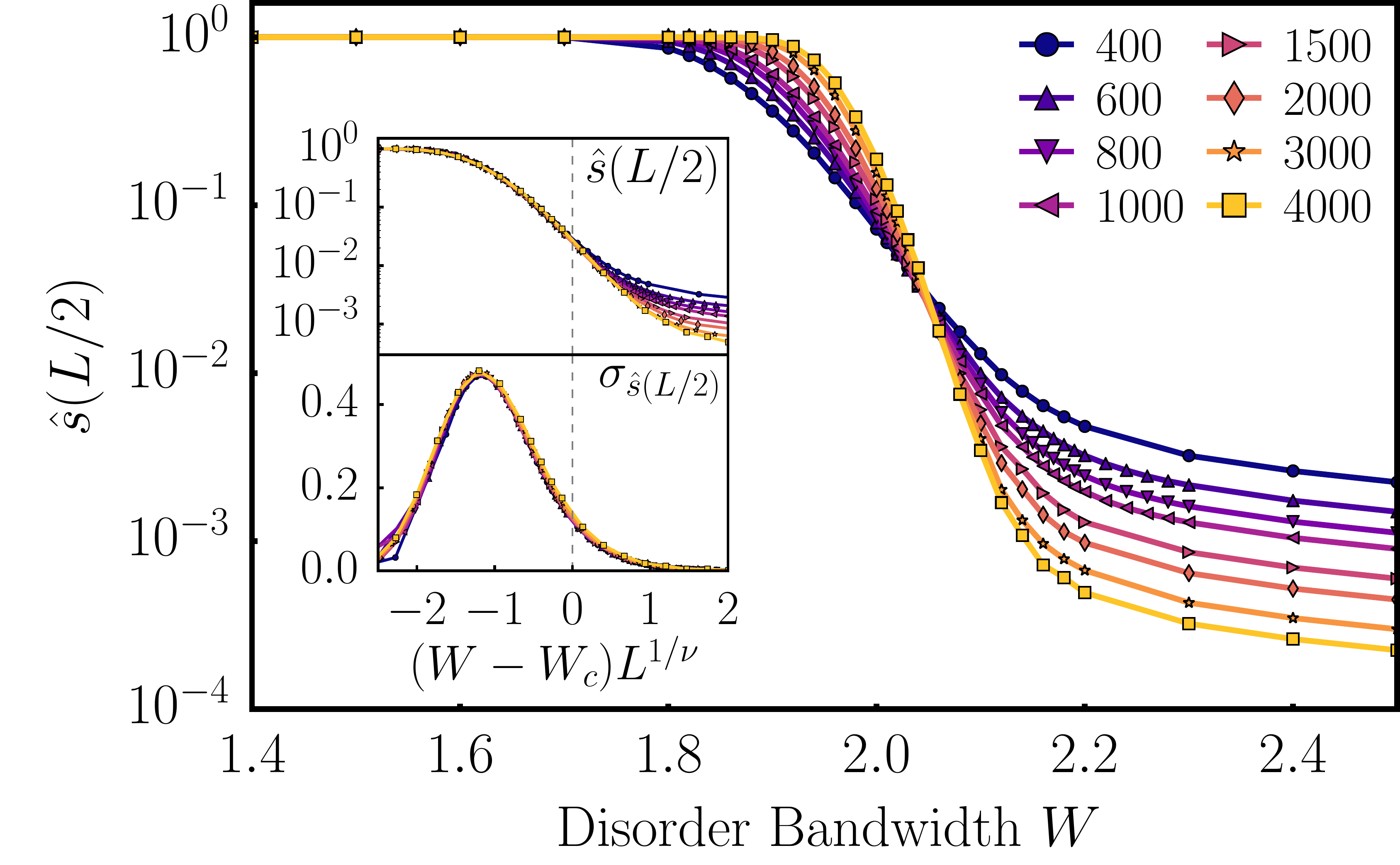}
\caption{
\label{fig:EntanglementCrossingL2} 
{\bf Universal scaling of bipartite entanglement -- } 
Normalized bipartite entanglement $\hat{s}(L/2)$ as a function of disorder bandwidth $W$ for different system sizes $L$.
Inset:~scaling collapse of $\hat{s}(L/2)$ [upper] and fluctuations $\sigma_{\hat{s}(L/2)}$ [lower], with $W_c = 2.05$ and $\nu = 3.2$.
Data with $L \leq 1000$ have $2.5\cdot 10^5$ or $10^6$ disorder realizations; those with $L \geq 1500$ have $10^5$.
Error-bars were calculated using Jackknife resampling, but are not shown when smaller than marker sizes.
}
\vspace{-0.2in}
\end{figure}

\header{Half-system Entanglement at Criticality} 
For each disorder realization, the RG produces a configuration of decoupled locally thermal clusters.
We calculate the entanglement of a subinterval by summing the thermal volume law contribution from each cluster spanning the interval boundaries. A cluster partitioned into $m$ and $n$ sites contributes $S_{m,n} = \min(m,n)s_\text{th}$.

Figure \ref{fig:EntanglementCrossingL2} depicts the normalized entanglement entropy, $\hat{s}(x) = \overline{S(x,L)} / x s_\text{th} $, for $x = L /2$, where $\overline{\(\dots\)}$ denotes averaging over disorder realizations and interval location. 
It shows the transition from a fully thermal system consisting of a single large cluster to the localized system made from many small clusters, indicated by curves of different $L$ crossing at critical disorder $W_c = 2.05\pm 0.01$.
The curves satisfy a scaling form $\hat{s} = f(\[W-W_{c}\] L^{1/\nu})$, with critical exponent $\nu = 3.2 \pm 0.3$ (Fig.~\ref{fig:EntanglementCrossingL2} upper inset).
This indicates the presence of a single diverging correlation length $\xi \approx |W-W_c|^{-\nu}$. 
A variety of observables give the same estimates of $W_c$ and $\nu$ and our extracted $\nu$ lies within error-bars of those obtained in \cite{Potter:2015aa,Vosk:2015aa}.
Notably, we find two distinct values of $\nu$ for average and typical correlation length exponents $\nu_\text{typ}\approx 2.1\pm 0.2$~\cite{SupMat}, consistent with a flow to infinite randomness. Together with the small value $\hat{s}$ at the crossing, this demonstrates that the transition is driven by rare thermal clusters separated by large MBL regions.

\begin{figure}[t]
\centering
\includegraphics[width=\columnwidth]{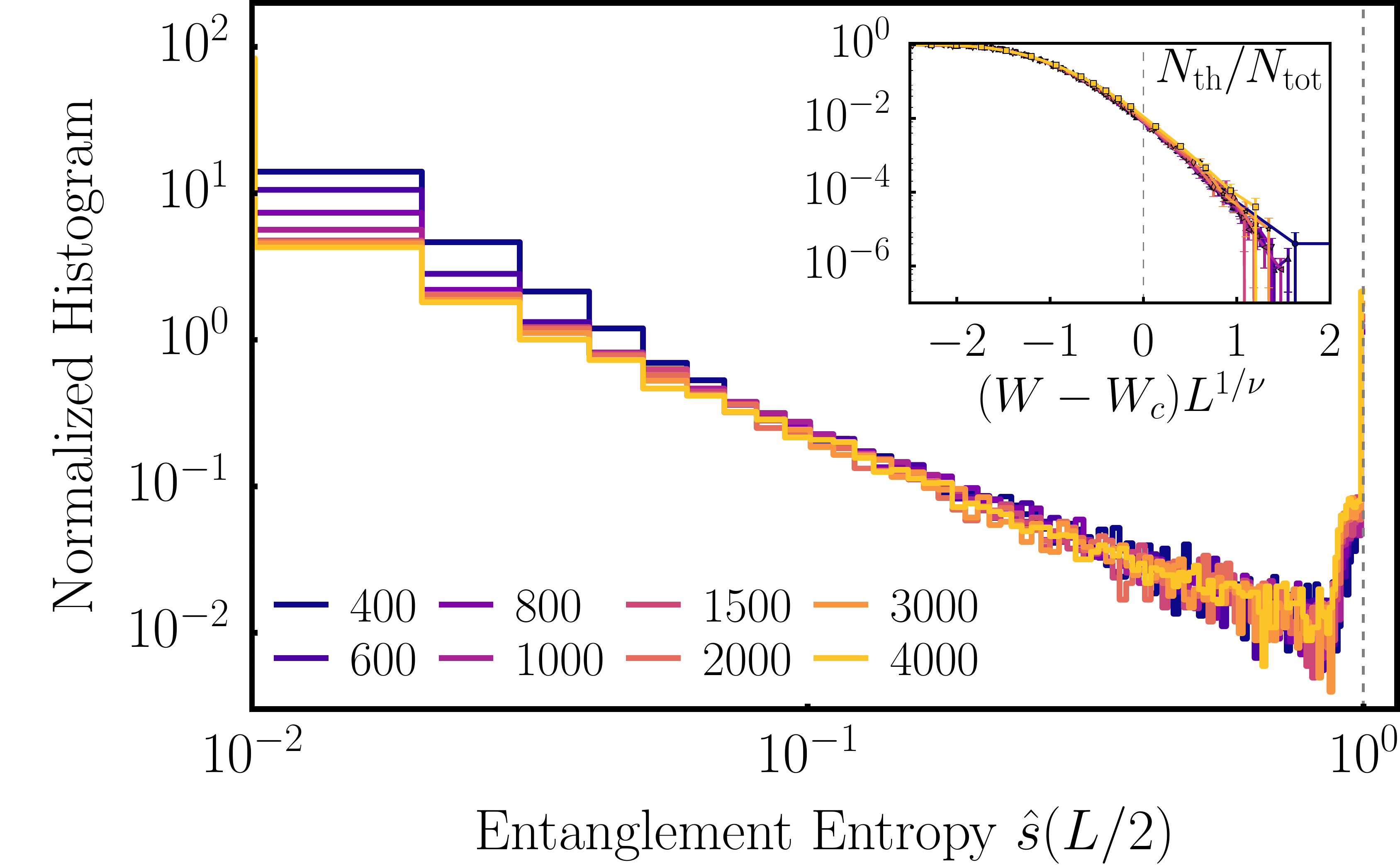}
\caption{\label{fig:EntanglementHistogramL2} 
{\bf Bipartite entanglement at criticality -- } 
Normalized histogram over 
disorder realizations of the bipartite entanglement entropy near criticality ($W = 2.04$), using 100 linearly spaced bins.
Inset:~scaling collapse of the fraction of fully thermalized 
configurations $N_{\mathrm{th}}/N_{\mathrm{tot}}$; error bars are 95\% confidence intervals expected for binomial distribution. 
}
\vspace{-0.2in}
\end{figure}

Figure~\ref{fig:EntanglementHistogramL2} shows the full histogram of entanglement over disorder realizations at $W_c$.
The distribution has a bimodal structure consisting of a power-law tail, $P(s)\approx s^{-\alpha}$ with $\alpha = 1.4 \pm 0.2$, fit over the interval $s\in [0.1,0.8]$, and a distinct sharp peak near the fully thermal value $s=1$. Away from criticality, the weight of the thermal peak scales like a universal function of $L/\xi$
(Fig.~\ref{fig:EntanglementHistogramL2} inset). Indications of a bimodal structure were observed in small-scale ED 
simulations~\cite{yu2016bimodal}. Our RG approach allows an extensive exploration of this structure.  

At criticality, the thermal peak gives a volume law contribution to the bipartite entanglement with a coefficient $a = (0.8 \pm 0.3) \cdot 10^{-2}$ far below the thermal value. The power-law component gives a universal sub-leading power-law contribution intermediate between area- and volume law,
\begin{align}
\overline{S(x=L/2,L,W=W_c)}\approx ax+bx^{1-\alpha}+\dots
\end{align}
These results differ from those of \cite{Vosk:2015aa}, whose proxy for half-system entanglement showed a smaller power-law ($P(s)\approx s^{-0.9}$) and lacked a thermal peak.

\header{Non-local influence of system size}
Consider next an infinite system slightly away from the critical point. Near a conventional continuous phase transition, observables (including entanglement) measured over distance $x$ exhibit critical behavior over an extended ``critical fan" $x\ll \xi$ extending across both sides of the transition. Moreover, they become independent of system size as $L \to \infty$, since critical fluctuations are determined by local physics. 
Entanglement at the MBLT departs dramatically from this conventional behavior, and instead shows a strong non-local dependence on system size, since an infinite thermal system can act as a bath for any finite subsystem, no matter its local properties. Hence, all subintervals of an infinite system must exhibit fully thermal entanglement $\hat{s}(x,L=\infty) = 1$ for $W<W_c$ and $L\gg\xi$~\cite{2014arXiv1405.1471G}. The conventional scaling picture would then suggest full thermal entanglement also on the MBL side ($W>W_c$) for $x\ll \xi$~\cite{2014arXiv1405.1471G}. 
Instead, ED simulations in~\cite{khemani2016critical} give evidence that this region actually has sub-thermal entanglement, consistent with the picture of~\cite{Potter:2015aa} that the critical regime mainly contains large MBL regions. Together with~\cite{2014arXiv1405.1471G}, this implies that the entanglement jumps discontinuously from fully thermal to sub-thermal across the MBLT for $L=\infty$~\cite{khemani2016critical}.

\begin{figure}[tb]
\centering
\includegraphics[width=\columnwidth]{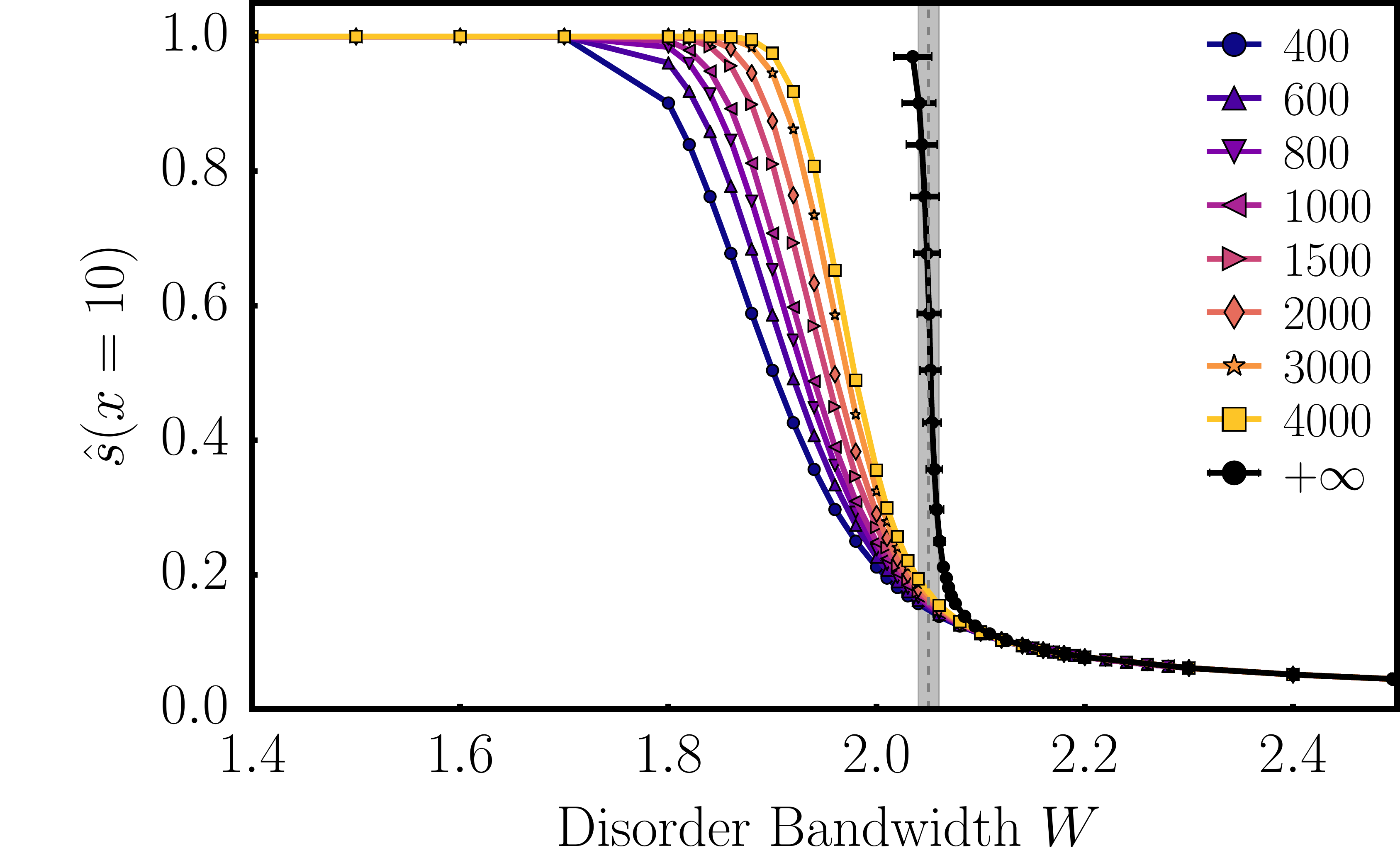}
\caption{\label{fig:EntanglementJump} 
{\bf  Infinite system entanglement -- } 
The normalized entanglement entropy for an interval $x=10$ develops a non-analytic step on the thermal side of the MBLT as $L\rightarrow \infty$.
Points labeled $+\infty$ are extrapolations in $L$, assuming the leading scaling form $\propto L^{1/\nu}$ along fixed $\hat{s}$. Cubic spline interpolation was used between data points. 
The error bars reflect the uncertainty in $\nu = 3.2 \pm 0.3$. The transition $W_{c} = 2.05 \pm 0.1$ is indicated by the the dashed line and gray shaded region.
}
\vspace{-0.2in}
\end{figure}

Our RG approach can directly demonstrate this predicted discontinuity by systematically extrapolating to the limit $L \rightarrow \infty$ with $x\ll \xi\ll L$.
Figure~\ref{fig:EntanglementJump} shows the normalized entanglement for a fixed interval $x = 10$ and various system sizes $L$. While one can never observe a true discontinuity in a finite size system, we observe a clear finite size flow towards a non-analytic jump with increasing $L$. Similar $L\rightarrow \infty$ extrapolations are obtained for all $x$.

\begin{figure}[tb]
\centering
\includegraphics[width=\columnwidth]{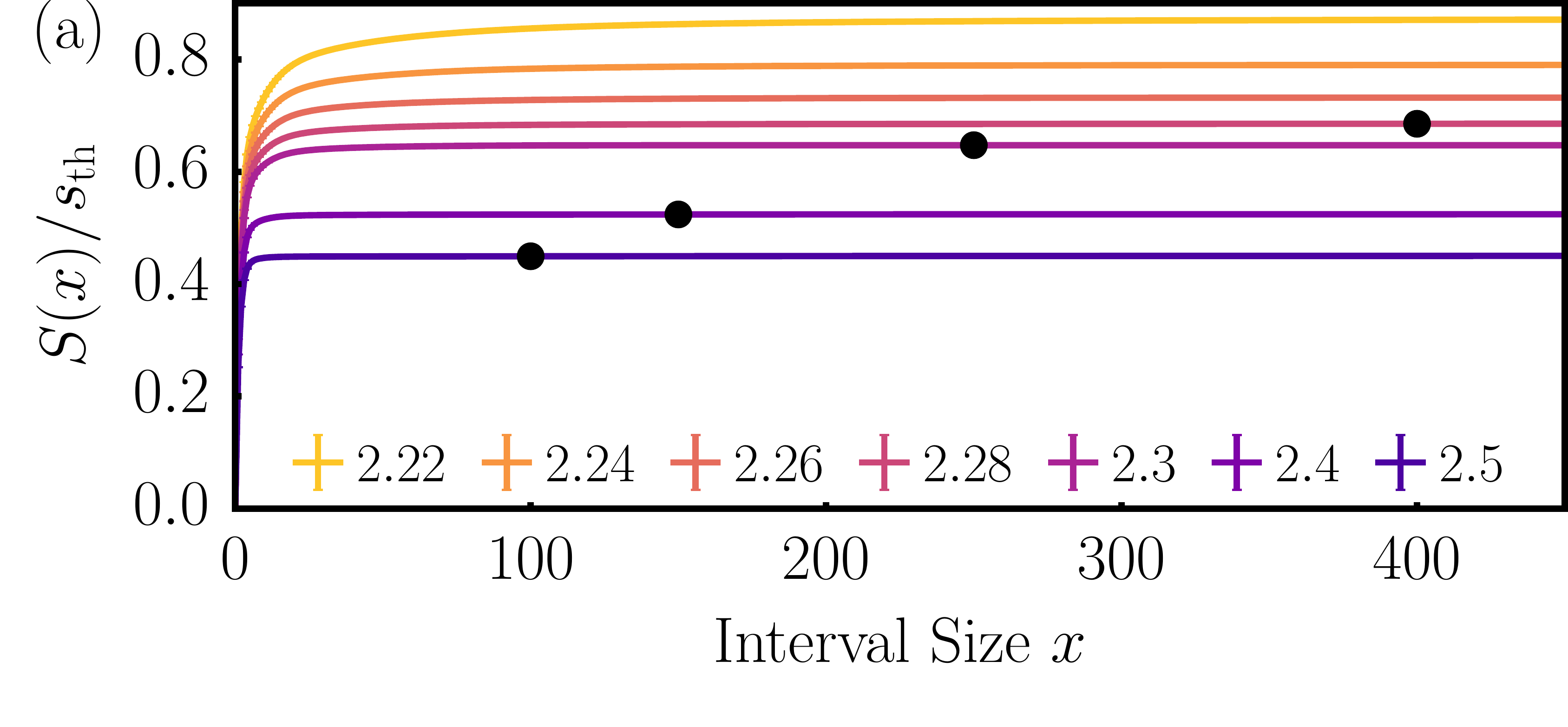}\\
\includegraphics[width=\columnwidth]{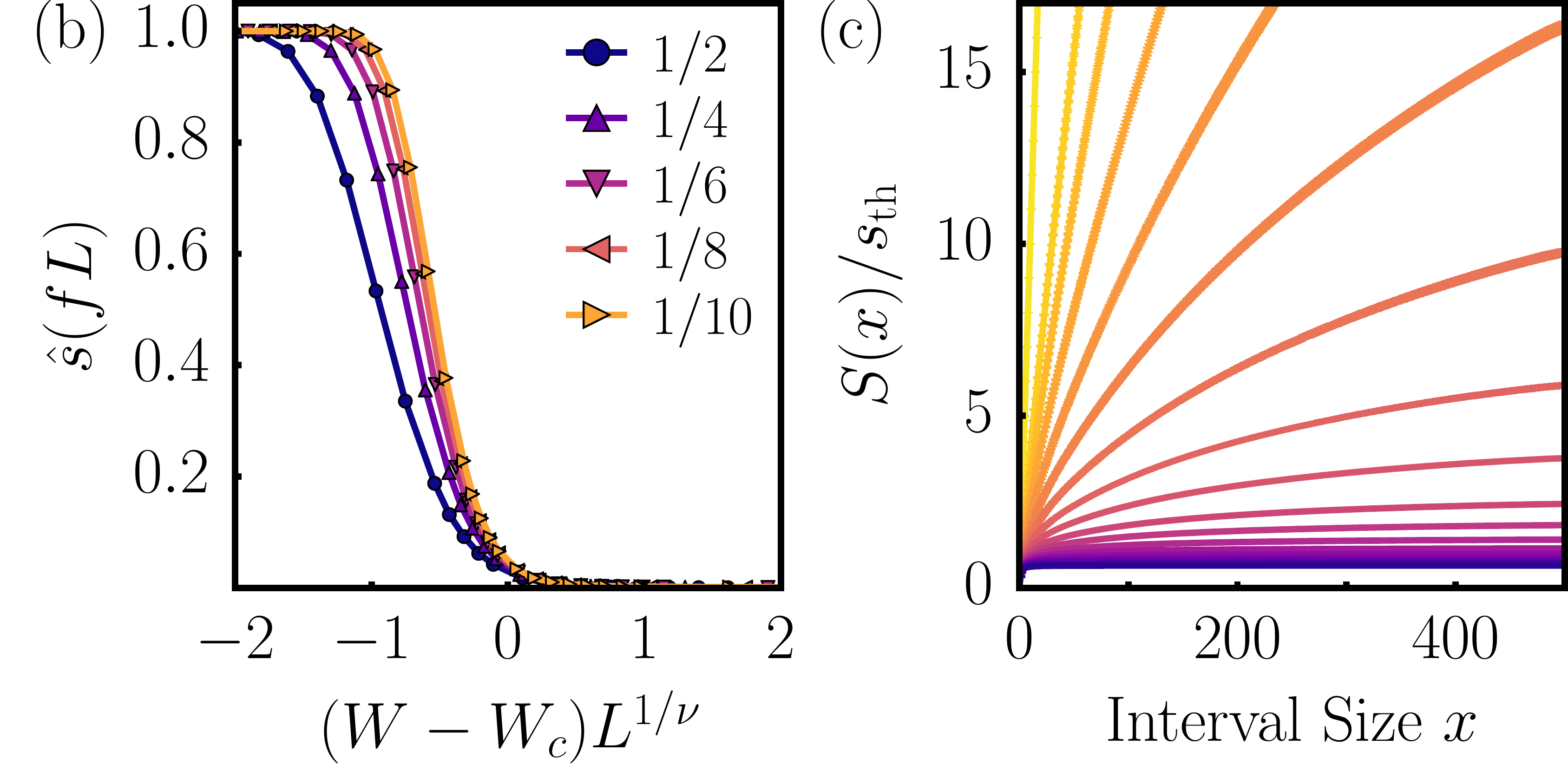}
\caption{\label{fig:S-x-dependence} 
{\bf  Entanglement finite-size crossover -- } 
(a)~Entanglement entropy as a function of interval size $x$ for system size $L = 1000$ and various 
$W>W_c$. The error bars correspond to the vertical thickness of the curves. The black points are lower bounds on the estimate of $\xi$ taken from the cluster size histograms.
(b)~Normalized entanglement entropy for $L = 4000$ for different fractions of system size $f = x / L$.
(c)~Plot as in (a) zoomed out for disorder values $W =$ 1.4 (yellow, linear volume law) and $1.96 \leq W \leq 2.3$ in steps of 0.02. 
}
\vspace{-0.25in}
\end{figure}

This discontinuous jump establishes that the entanglement on the MBL side is sub-thermal for all $x$. However, many functional forms are consistent with this requirement. Unlike the thermal behavior for $L=\infty$ and $W<W_c$, which follows from analytic constraints \cite{2014arXiv1405.1471G}, determining the entanglement scaling for $W>W_c$ requires a three-fold hierarchy of scales $1\ll x\ll \xi \ll L$ (Fig.~\ref{fig:S-x-dependence}). This necessitates large systems with at least $\mathcal{O}(10^3)$ sites, making our RG approach uniquely suited to address this question. 

Having an objective measure of the correlation length $\xi$ is vital to identify the desired scaling regime and separate it from the distinct crossover behavior when $\xi \approx L$.
To this end, we examine the distribution of cluster sizes, which exhibit power-law decay up to a scale that we identify as $\xi$, beyond which they decay exponentially~\cite{SupMat}. For $L\gg \xi$, the entanglement curves show a small non-universal rise over $x \lesssim1-10$ and then remain perfectly flat as $x$ crosses through $\xi$, indicating that the entanglement follows a pure area-law everywhere on the MBL side of the transition, even for $x\ll \xi$. 

{The absence of scaling on the MBL side is particular to the disorder averaged entanglement, for which critical fluctuations affect only subleading terms that vanish for large $L$. Other observables, like higher moments of entanglement can exhibit universal power-law singularities as $W\rightarrow W_c^+$. We also note that the discontinuous behavior of entanglement for $L\rightarrow \infty$ is special to static eigenstate properties (equivalently, infinite time averaged quantities). In contrast, due to the logarithmic causal-cone for dynamics at the MBLT~\cite{Vosk:2015aa,Potter:2015aa}, dynamical measurements on timescales $\log t\ll L$ are insensitive to the system size, and will exhibit a more conventional critical scaling fan. }

\header{Full scaling form of $S(x,L,W)$}
For infinite systems, we have seen that the entanglement jumps discontinuously at the MBLT. For finite $L$, this jump becomes a smooth crossover. What universal data can we extract from this crossover? The entanglement is itself generically not a scaling variable. In addition to   non-universal, sub-leading terms, different parts of the entanglement may be universal for different critical points; identifying an appropriate scaling form is not straightforward. For example, in one-dimensional conformal field theories one needs to consider
${\d  S}/{\d \log x}$ in the  limit  $x,L\gg 1$~\cite{Myers2012,Liu2013}. 

By performing scaling collapses of $S(x,L)$ for fixed ${x}/{L}$ and various $W$~\cite{SupMat}, we find evidence that the volume law coefficient is a universal scaling function
\begin{align}\label{eq:univVolLaw}
\hat{s}(x,L) = \frac{\overline{S(x,L)}}{xs_\text{th}} = \mathcal{A}\(\frac{x}{\xi},\frac{L}{\xi}, \sgn\delta W\) + (\dots).
\end{align}
Here  $(\dots)$ indicates sub-leading corrections in  $x$ and $L$ that vanish in the scaling limit $x,L\gg 1$. The scaled form as the function of the variables $x/L, L / \xi \sgn \delta W$ is shown in Fig.~\ref{fig:S-x-dependence}b.  At finite $L$, the above scaling form with a single universal exponent $\nu$ is relatively conventional. The large $L$ limit, however, is different from the scaling of conventional correlation functions. The non-local system size dependence shown above, implies that in the limit $L / \xi \to \infty$, $\mathcal{A}$ depends only on $\sgn \delta W$; there is absolutely no dependence on $x / \xi$. The striking discrepancy in scaling highlights the unusual and asymmetric nature of thermalization and the MBL transition.

\begin{acknowledgments}
\vspace{4pt}\noindent{\it Acknowledgements -- }
We thank D.~Huse and T.~Grover for insightful discussions and especially S.A.~Parameswaran for collaboration on previous work. We acknowledge the Texas Advanced Computing Center (TACC) at the University of Texas at Austin for computational resources and LBNL Quantum Materials program for support (R.V.).
\end{acknowledgments}

\bibliography{mbl-biblio}

\begin{thebibliography}{58}%
\makeatletter
\providecommand \@ifxundefined [1]{%
 \@ifx{#1\undefined}
}%
\providecommand \@ifnum [1]{%
 \ifnum #1\expandafter \@firstoftwo
 \else \expandafter \@secondoftwo
 \fi
}%
\providecommand \@ifx [1]{%
 \ifx #1\expandafter \@firstoftwo
 \else \expandafter \@secondoftwo
 \fi
}%
\providecommand \natexlab [1]{#1}%
\providecommand \enquote  [1]{``#1''}%
\providecommand \bibnamefont  [1]{#1}%
\providecommand \bibfnamefont [1]{#1}%
\providecommand \citenamefont [1]{#1}%
\providecommand \href@noop [0]{\@secondoftwo}%
\providecommand \href [0]{\begingroup \@sanitize@url \@href}%
\providecommand \@href[1]{\@@startlink{#1}\@@href}%
\providecommand \@@href[1]{\endgroup#1\@@endlink}%
\providecommand \@sanitize@url [0]{\catcode `\\12\catcode `\$12\catcode
  `\&12\catcode `\#12\catcode `\^12\catcode `\_12\catcode `\%12\relax}%
\providecommand \@@startlink[1]{}%
\providecommand \@@endlink[0]{}%
\providecommand \url  [0]{\begingroup\@sanitize@url \@url }%
\providecommand \@url [1]{\endgroup\@href {#1}{\urlprefix }}%
\providecommand \urlprefix  [0]{URL }%
\providecommand \Eprint [0]{\href }%
\providecommand \doibase [0]{http://dx.doi.org/}%
\providecommand \selectlanguage [0]{\@gobble}%
\providecommand \bibinfo  [0]{\@secondoftwo}%
\providecommand \bibfield  [0]{\@secondoftwo}%
\providecommand \translation [1]{[#1]}%
\providecommand \BibitemOpen [0]{}%
\providecommand \bibitemStop [0]{}%
\providecommand \bibitemNoStop [0]{.\EOS\space}%
\providecommand \EOS [0]{\spacefactor3000\relax}%
\providecommand \BibitemShut  [1]{\csname bibitem#1\endcsname}%
\let\auto@bib@innerbib\@empty
%</preamble>
\bibitem [{\citenamefont {Bloch}\ \emph {et~al.}(2008)\citenamefont {Bloch},
  \citenamefont {Dalibard},\ and\ \citenamefont {Zwerger}}]{bloch_review}%
  \BibitemOpen
  \bibfield  {author} {\bibinfo {author} {\bibfnamefont {I.}~\bibnamefont
  {Bloch}}, \bibinfo {author} {\bibfnamefont {J.}~\bibnamefont {Dalibard}}, \
  and\ \bibinfo {author} {\bibfnamefont {W.}~\bibnamefont {Zwerger}},\ }\href
  {\doibase 10.1103/RevModPhys.80.885} {\bibfield  {journal} {\bibinfo
  {journal} {Rev. Mod. Phys.}\ }\textbf {\bibinfo {volume} {80}},\ \bibinfo
  {pages} {885} (\bibinfo {year} {2008})}\BibitemShut {NoStop}%
\bibitem [{\citenamefont {Schreiber}\ \emph {et~al.}(2015)\citenamefont
  {Schreiber}, \citenamefont {Hodgman}, \citenamefont {Bordia}, \citenamefont
  {L{\"u}schen}, \citenamefont {Fischer}, \citenamefont {Vosk}, \citenamefont
  {Altman}, \citenamefont {Schneider},\ and\ \citenamefont
  {Bloch}}]{Schreiber842}%
  \BibitemOpen
  \bibfield  {author} {\bibinfo {author} {\bibfnamefont {M.}~\bibnamefont
  {Schreiber}}, \bibinfo {author} {\bibfnamefont {S.~S.}\ \bibnamefont
  {Hodgman}}, \bibinfo {author} {\bibfnamefont {P.}~\bibnamefont {Bordia}},
  \bibinfo {author} {\bibfnamefont {H.~P.}\ \bibnamefont {L{\"u}schen}},
  \bibinfo {author} {\bibfnamefont {M.~H.}\ \bibnamefont {Fischer}}, \bibinfo
  {author} {\bibfnamefont {R.}~\bibnamefont {Vosk}}, \bibinfo {author}
  {\bibfnamefont {E.}~\bibnamefont {Altman}}, \bibinfo {author} {\bibfnamefont
  {U.}~\bibnamefont {Schneider}}, \ and\ \bibinfo {author} {\bibfnamefont
  {I.}~\bibnamefont {Bloch}},\ }\href {\doibase 10.1126/science.aaa7432}
  {\bibfield  {journal} {\bibinfo  {journal} {Science}\ }\textbf {\bibinfo
  {volume} {349}},\ \bibinfo {pages} {842} (\bibinfo {year}
  {2015})}\BibitemShut {NoStop}%
\bibitem [{\citenamefont {Choi}\ \emph {et~al.}(2016)\citenamefont {Choi},
  \citenamefont {Hild}, \citenamefont {Zeiher}, \citenamefont {Schau{\ss}},
  \citenamefont {Rubio-Abadal}, \citenamefont {Yefsah}, \citenamefont
  {Khemani}, \citenamefont {Huse}, \citenamefont {Bloch},\ and\ \citenamefont
  {Gross}}]{choi2016exploring}%
  \BibitemOpen
  \bibfield  {author} {\bibinfo {author} {\bibfnamefont {J.-y.}\ \bibnamefont
  {Choi}}, \bibinfo {author} {\bibfnamefont {S.}~\bibnamefont {Hild}}, \bibinfo
  {author} {\bibfnamefont {J.}~\bibnamefont {Zeiher}}, \bibinfo {author}
  {\bibfnamefont {P.}~\bibnamefont {Schau{\ss}}}, \bibinfo {author}
  {\bibfnamefont {A.}~\bibnamefont {Rubio-Abadal}}, \bibinfo {author}
  {\bibfnamefont {T.}~\bibnamefont {Yefsah}}, \bibinfo {author} {\bibfnamefont
  {V.}~\bibnamefont {Khemani}}, \bibinfo {author} {\bibfnamefont {D.~A.}\
  \bibnamefont {Huse}}, \bibinfo {author} {\bibfnamefont {I.}~\bibnamefont
  {Bloch}}, \ and\ \bibinfo {author} {\bibfnamefont {C.}~\bibnamefont
  {Gross}},\ }\href {\doibase 10.1126/science.aaf8834} {\bibfield  {journal}
  {\bibinfo  {journal} {Science}\ }\textbf {\bibinfo {volume} {352}},\ \bibinfo
  {pages} {1547} (\bibinfo {year} {2016})}\BibitemShut {NoStop}%
\bibitem [{\citenamefont {{L{\"u}schen}}\ \emph {et~al.}(2016)\citenamefont
  {{L{\"u}schen}}, \citenamefont {{Bordia}}, \citenamefont {{Scherg}},
  \citenamefont {{Alet}}, \citenamefont {{Altman}}, \citenamefont
  {{Schneider}},\ and\ \citenamefont {{Bloch}}}]{2016arXiv161207173L}%
  \BibitemOpen
  \bibfield  {author} {\bibinfo {author} {\bibfnamefont {H.~P.}\ \bibnamefont
  {{L{\"u}schen}}}, \bibinfo {author} {\bibfnamefont {P.}~\bibnamefont
  {{Bordia}}}, \bibinfo {author} {\bibfnamefont {S.}~\bibnamefont {{Scherg}}},
  \bibinfo {author} {\bibfnamefont {F.}~\bibnamefont {{Alet}}}, \bibinfo
  {author} {\bibfnamefont {E.}~\bibnamefont {{Altman}}}, \bibinfo {author}
  {\bibfnamefont {U.}~\bibnamefont {{Schneider}}}, \ and\ \bibinfo {author}
  {\bibfnamefont {I.}~\bibnamefont {{Bloch}}},\ }\href@noop {} {\bibfield
  {journal} {\bibinfo  {journal} {ArXiv e-prints}\ } (\bibinfo {year}
  {2016})},\ \Eprint {http://arxiv.org/abs/1612.07173} {arXiv:1612.07173
  [cond-mat.quant-gas]} \BibitemShut {NoStop}%
\bibitem [{\citenamefont {Smith}\ \emph {et~al.}(2016)\citenamefont {Smith},
  \citenamefont {Lee}, \citenamefont {Richerme}, \citenamefont {Neyenhuis},
  \citenamefont {Hess}, \citenamefont {Hauke}, \citenamefont {Heyl},
  \citenamefont {Huse},\ and\ \citenamefont {Monroe}}]{Smith2016aa}%
  \BibitemOpen
  \bibfield  {author} {\bibinfo {author} {\bibfnamefont {J.}~\bibnamefont
  {Smith}}, \bibinfo {author} {\bibfnamefont {A.}~\bibnamefont {Lee}}, \bibinfo
  {author} {\bibfnamefont {P.}~\bibnamefont {Richerme}}, \bibinfo {author}
  {\bibfnamefont {B.}~\bibnamefont {Neyenhuis}}, \bibinfo {author}
  {\bibfnamefont {P.~W.}\ \bibnamefont {Hess}}, \bibinfo {author}
  {\bibfnamefont {P.}~\bibnamefont {Hauke}}, \bibinfo {author} {\bibfnamefont
  {M.}~\bibnamefont {Heyl}}, \bibinfo {author} {\bibfnamefont {D.~A.}\
  \bibnamefont {Huse}}, \ and\ \bibinfo {author} {\bibfnamefont
  {C.}~\bibnamefont {Monroe}},\ }\href {http://dx.doi.org/10.1038/nphys3783}
  {\bibfield  {journal} {\bibinfo  {journal} {Nat. Phys.}\ }\textbf {\bibinfo
  {volume} {12}},\ \bibinfo {pages} {907} (\bibinfo {year} {2016})}\BibitemShut
  {NoStop}%
\bibitem [{\citenamefont {Zhang}\ \emph {et~al.}(2017)\citenamefont {Zhang},
  \citenamefont {Hess}, \citenamefont {Kyprianidis}, \citenamefont {Becker},
  \citenamefont {Lee}, \citenamefont {Smith}, \citenamefont {Pagano},
  \citenamefont {Potirniche}, \citenamefont {Potter}, \citenamefont
  {Vishwanath}, \citenamefont {Yao},\ and\ \citenamefont
  {Monroe}}]{2016arXiv160908684Z}%
  \BibitemOpen
  \bibfield  {author} {\bibinfo {author} {\bibfnamefont {J.}~\bibnamefont
  {Zhang}}, \bibinfo {author} {\bibfnamefont {P.~W.}\ \bibnamefont {Hess}},
  \bibinfo {author} {\bibfnamefont {A.}~\bibnamefont {Kyprianidis}}, \bibinfo
  {author} {\bibfnamefont {P.}~\bibnamefont {Becker}}, \bibinfo {author}
  {\bibfnamefont {A.}~\bibnamefont {Lee}}, \bibinfo {author} {\bibfnamefont
  {J.}~\bibnamefont {Smith}}, \bibinfo {author} {\bibfnamefont
  {G.}~\bibnamefont {Pagano}}, \bibinfo {author} {\bibfnamefont {I.~D.}\
  \bibnamefont {Potirniche}}, \bibinfo {author} {\bibfnamefont {A.~C.}\
  \bibnamefont {Potter}}, \bibinfo {author} {\bibfnamefont {A.}~\bibnamefont
  {Vishwanath}}, \bibinfo {author} {\bibfnamefont {N.~Y.}\ \bibnamefont {Yao}},
  \ and\ \bibinfo {author} {\bibfnamefont {C.}~\bibnamefont {Monroe}},\ }\href
  {http://dx.doi.org/10.1038/nature21413} {\bibfield  {journal} {\bibinfo
  {journal} {Nature}\ }\textbf {\bibinfo {volume} {543}},\ \bibinfo {pages}
  {217} (\bibinfo {year} {2017})}\BibitemShut {NoStop}%
\bibitem [{\citenamefont {Choi}\ \emph {et~al.}(2017)\citenamefont {Choi},
  \citenamefont {Choi}, \citenamefont {Landig}, \citenamefont {Kucsko},
  \citenamefont {Zhou}, \citenamefont {Isoya}, \citenamefont {Jelezko},
  \citenamefont {Onoda}, \citenamefont {Sumiya}, \citenamefont {Khemani},
  \citenamefont {von Keyserlingk}, \citenamefont {Yao}, \citenamefont
  {Demler},\ and\ \citenamefont {Lukin}}]{2016arXiv161008057C}%
  \BibitemOpen
  \bibfield  {author} {\bibinfo {author} {\bibfnamefont {S.}~\bibnamefont
  {Choi}}, \bibinfo {author} {\bibfnamefont {J.}~\bibnamefont {Choi}}, \bibinfo
  {author} {\bibfnamefont {R.}~\bibnamefont {Landig}}, \bibinfo {author}
  {\bibfnamefont {G.}~\bibnamefont {Kucsko}}, \bibinfo {author} {\bibfnamefont
  {H.}~\bibnamefont {Zhou}}, \bibinfo {author} {\bibfnamefont {J.}~\bibnamefont
  {Isoya}}, \bibinfo {author} {\bibfnamefont {F.}~\bibnamefont {Jelezko}},
  \bibinfo {author} {\bibfnamefont {S.}~\bibnamefont {Onoda}}, \bibinfo
  {author} {\bibfnamefont {H.}~\bibnamefont {Sumiya}}, \bibinfo {author}
  {\bibfnamefont {V.}~\bibnamefont {Khemani}}, \bibinfo {author} {\bibfnamefont
  {C.}~\bibnamefont {von Keyserlingk}}, \bibinfo {author} {\bibfnamefont
  {N.~Y.}\ \bibnamefont {Yao}}, \bibinfo {author} {\bibfnamefont
  {E.}~\bibnamefont {Demler}}, \ and\ \bibinfo {author} {\bibfnamefont {M.~D.}\
  \bibnamefont {Lukin}},\ }\href {http://dx.doi.org/10.1038/nature21426}
  {\bibfield  {journal} {\bibinfo  {journal} {Nature}\ }\textbf {\bibinfo
  {volume} {543}},\ \bibinfo {pages} {221} (\bibinfo {year}
  {2017})}\BibitemShut {NoStop}%
\bibitem [{\citenamefont {{Wei}}\ \emph {et~al.}(2016)\citenamefont {{Wei}},
  \citenamefont {{Ramanathan}},\ and\ \citenamefont
  {{Cappellaro}}}]{2016arXiv161205249W}%
  \BibitemOpen
  \bibfield  {author} {\bibinfo {author} {\bibfnamefont {K.~X.}\ \bibnamefont
  {{Wei}}}, \bibinfo {author} {\bibfnamefont {C.}~\bibnamefont {{Ramanathan}}},
  \ and\ \bibinfo {author} {\bibfnamefont {P.}~\bibnamefont {{Cappellaro}}},\
  }\href@noop {} {\bibfield  {journal} {\bibinfo  {journal} {ArXiv e-prints}\ }
  (\bibinfo {year} {2016})},\ \Eprint {http://arxiv.org/abs/1612.05249}
  {arXiv:1612.05249 [cond-mat.dis-nn]} \BibitemShut {NoStop}%
\bibitem [{\citenamefont {Srednicki}(1994)}]{PhysRevE.50.888}%
  \BibitemOpen
  \bibfield  {author} {\bibinfo {author} {\bibfnamefont {M.}~\bibnamefont
  {Srednicki}},\ }\href {\doibase 10.1103/PhysRevE.50.888} {\bibfield
  {journal} {\bibinfo  {journal} {Phys. Rev. E}\ }\textbf {\bibinfo {volume}
  {50}},\ \bibinfo {pages} {888} (\bibinfo {year} {1994})}\BibitemShut
  {NoStop}%
\bibitem [{\citenamefont {Deutsch}(1991)}]{PhysRevA.43.2046}%
  \BibitemOpen
  \bibfield  {author} {\bibinfo {author} {\bibfnamefont {J.~M.}\ \bibnamefont
  {Deutsch}},\ }\href {\doibase 10.1103/PhysRevA.43.2046} {\bibfield  {journal}
  {\bibinfo  {journal} {Phys. Rev. A}\ }\textbf {\bibinfo {volume} {43}},\
  \bibinfo {pages} {2046} (\bibinfo {year} {1991})}\BibitemShut {NoStop}%
\bibitem [{\citenamefont {Anderson}(1958)}]{PhysRev.109.1492}%
  \BibitemOpen
  \bibfield  {author} {\bibinfo {author} {\bibfnamefont {P.~W.}\ \bibnamefont
  {Anderson}},\ }\href {\doibase 10.1103/PhysRev.109.1492} {\bibfield
  {journal} {\bibinfo  {journal} {Phys. Rev.}\ }\textbf {\bibinfo {volume}
  {109}},\ \bibinfo {pages} {1492} (\bibinfo {year} {1958})}\BibitemShut
  {NoStop}%
\bibitem [{\citenamefont {Fleishman}\ and\ \citenamefont
  {Anderson}(1980)}]{FleishmanAnderson}%
  \BibitemOpen
  \bibfield  {author} {\bibinfo {author} {\bibfnamefont {L.}~\bibnamefont
  {Fleishman}}\ and\ \bibinfo {author} {\bibfnamefont {P.~W.}\ \bibnamefont
  {Anderson}},\ }\href {\doibase 10.1103/PhysRevB.21.2366} {\bibfield
  {journal} {\bibinfo  {journal} {Phys. Rev. B}\ }\textbf {\bibinfo {volume}
  {21}},\ \bibinfo {pages} {2366} (\bibinfo {year} {1980})}\BibitemShut
  {NoStop}%
\bibitem [{\citenamefont {Gornyi}\ \emph {et~al.}(2005)\citenamefont {Gornyi},
  \citenamefont {Mirlin},\ and\ \citenamefont {Polyakov}}]{Gornyi}%
  \BibitemOpen
  \bibfield  {author} {\bibinfo {author} {\bibfnamefont {I.~V.}\ \bibnamefont
  {Gornyi}}, \bibinfo {author} {\bibfnamefont {A.~D.}\ \bibnamefont {Mirlin}},
  \ and\ \bibinfo {author} {\bibfnamefont {D.~G.}\ \bibnamefont {Polyakov}},\
  }\href {\doibase 10.1103/PhysRevLett.95.206603} {\bibfield  {journal}
  {\bibinfo  {journal} {Phys. Rev. Lett.}\ }\textbf {\bibinfo {volume} {95}},\
  \bibinfo {pages} {206603} (\bibinfo {year} {2005})}\BibitemShut {NoStop}%
\bibitem [{\citenamefont {Basko}\ \emph {et~al.}(2006)\citenamefont {Basko},
  \citenamefont {Aleiner},\ and\ \citenamefont {Altshuler}}]{BAA}%
  \BibitemOpen
  \bibfield  {author} {\bibinfo {author} {\bibfnamefont {D.}~\bibnamefont
  {Basko}}, \bibinfo {author} {\bibfnamefont {I.}~\bibnamefont {Aleiner}}, \
  and\ \bibinfo {author} {\bibfnamefont {B.}~\bibnamefont {Altshuler}},\ }\href
  {\doibase 10.1016/j.aop.2005.11.014} {\bibfield  {journal} {\bibinfo
  {journal} {Ann. Phys. (NY)}\ }\textbf {\bibinfo {volume} {321}},\ \bibinfo
  {pages} {1126 } (\bibinfo {year} {2006})}\BibitemShut {NoStop}%
\bibitem [{\citenamefont {Oganesyan}\ and\ \citenamefont
  {Huse}(2007)}]{PhysRevB.75.155111}%
  \BibitemOpen
  \bibfield  {author} {\bibinfo {author} {\bibfnamefont {V.}~\bibnamefont
  {Oganesyan}}\ and\ \bibinfo {author} {\bibfnamefont {D.~A.}\ \bibnamefont
  {Huse}},\ }\href {\doibase 10.1103/PhysRevB.75.155111} {\bibfield  {journal}
  {\bibinfo  {journal} {Phys. Rev. B}\ }\textbf {\bibinfo {volume} {75}},\
  \bibinfo {pages} {155111} (\bibinfo {year} {2007})}\BibitemShut {NoStop}%
\bibitem [{\citenamefont {Pal}\ and\ \citenamefont {Huse}(2010)}]{PalHuse}%
  \BibitemOpen
  \bibfield  {author} {\bibinfo {author} {\bibfnamefont {A.}~\bibnamefont
  {Pal}}\ and\ \bibinfo {author} {\bibfnamefont {D.~A.}\ \bibnamefont {Huse}},\
  }\href {\doibase 10.1103/PhysRevB.82.174411} {\bibfield  {journal} {\bibinfo
  {journal} {Phys. Rev. B}\ }\textbf {\bibinfo {volume} {82}},\ \bibinfo
  {pages} {174411} (\bibinfo {year} {2010})}\BibitemShut {NoStop}%
\bibitem [{\citenamefont {Bauer}\ and\ \citenamefont
  {Nayak}(2013)}]{BauerNayak}%
  \BibitemOpen
  \bibfield  {author} {\bibinfo {author} {\bibfnamefont {B.}~\bibnamefont
  {Bauer}}\ and\ \bibinfo {author} {\bibfnamefont {C.}~\bibnamefont {Nayak}},\
  }\href {http://stacks.iop.org/1742-5468/2013/i=09/a=P09005} {\bibfield
  {journal} {\bibinfo  {journal} {Journal of Statistical Mechanics: Theory and
  Experiment}\ }\textbf {\bibinfo {volume} {2013}},\ \bibinfo {pages} {P09005}
  (\bibinfo {year} {2013})}\BibitemShut {NoStop}%
\bibitem [{\citenamefont {Nandkishore}\ and\ \citenamefont
  {Huse}(2015)}]{2014arXiv1404.0686N}%
  \BibitemOpen
  \bibfield  {author} {\bibinfo {author} {\bibfnamefont {R.}~\bibnamefont
  {Nandkishore}}\ and\ \bibinfo {author} {\bibfnamefont {D.~A.}\ \bibnamefont
  {Huse}},\ }\href {\doibase 10.1146/annurev-conmatphys-031214-014726}
  {\bibfield  {journal} {\bibinfo  {journal} {Annual Review of Condensed Matter
  Physics}\ }\textbf {\bibinfo {volume} {6}},\ \bibinfo {pages} {15} (\bibinfo
  {year} {2015})}\BibitemShut {NoStop}%
\bibitem [{\citenamefont {Altman}\ and\ \citenamefont
  {Vosk}(2015)}]{Altman:2015aa}%
  \BibitemOpen
  \bibfield  {author} {\bibinfo {author} {\bibfnamefont {E.}~\bibnamefont
  {Altman}}\ and\ \bibinfo {author} {\bibfnamefont {R.}~\bibnamefont {Vosk}},\
  }\href@noop {} {\bibfield  {journal} {\bibinfo  {journal} {Annu. Rev.
  Condens. Matter Phys.}\ }\textbf {\bibinfo {volume} {6}},\ \bibinfo {pages}
  {383} (\bibinfo {year} {2015})}\BibitemShut {NoStop}%
\bibitem [{\citenamefont {Vasseur}\ and\ \citenamefont
  {Moore}(2016)}]{1742-5468-2016-6-064010}%
  \BibitemOpen
  \bibfield  {author} {\bibinfo {author} {\bibfnamefont {R.}~\bibnamefont
  {Vasseur}}\ and\ \bibinfo {author} {\bibfnamefont {J.~E.}\ \bibnamefont
  {Moore}},\ }\href {http://stacks.iop.org/1742-5468/2016/i=6/a=064010}
  {\bibfield  {journal} {\bibinfo  {journal} {Journal of Statistical Mechanics:
  Theory and Experiment}\ }\textbf {\bibinfo {volume} {2016}},\ \bibinfo
  {pages} {064010} (\bibinfo {year} {2016})}\BibitemShut {NoStop}%
\bibitem [{\citenamefont {Znidaric}\ \emph {et~al.}(2008)\citenamefont
  {Znidaric}, \citenamefont {Prosen},\ and\ \citenamefont
  {Prelovsek}}]{PhysRevB.77.064426}%
  \BibitemOpen
  \bibfield  {author} {\bibinfo {author} {\bibfnamefont {M.}~\bibnamefont
  {Znidaric}}, \bibinfo {author} {\bibfnamefont {T.}~\bibnamefont {Prosen}}, \
  and\ \bibinfo {author} {\bibfnamefont {P.}~\bibnamefont {Prelovsek}},\ }\href
  {\doibase 10.1103/PhysRevB.77.064426} {\bibfield  {journal} {\bibinfo
  {journal} {Phys. Rev. B}\ }\textbf {\bibinfo {volume} {77}},\ \bibinfo
  {pages} {064426} (\bibinfo {year} {2008})}\BibitemShut {NoStop}%
\bibitem [{\citenamefont {Bardarson}\ \emph {et~al.}(2012)\citenamefont
  {Bardarson}, \citenamefont {Pollmann},\ and\ \citenamefont
  {Moore}}]{PhysRevLett.109.017202}%
  \BibitemOpen
  \bibfield  {author} {\bibinfo {author} {\bibfnamefont {J.~H.}\ \bibnamefont
  {Bardarson}}, \bibinfo {author} {\bibfnamefont {F.}~\bibnamefont {Pollmann}},
  \ and\ \bibinfo {author} {\bibfnamefont {J.~E.}\ \bibnamefont {Moore}},\
  }\href {\doibase 10.1103/PhysRevLett.109.017202} {\bibfield  {journal}
  {\bibinfo  {journal} {Phys. Rev. Lett.}\ }\textbf {\bibinfo {volume} {109}},\
  \bibinfo {pages} {017202} (\bibinfo {year} {2012})}\BibitemShut {NoStop}%
\bibitem [{\citenamefont {Serbyn}\ \emph
  {et~al.}(2013{\natexlab{a}})\citenamefont {Serbyn}, \citenamefont
  {Papi\ifmmode~\acute{c}\else \'{c}\fi{}},\ and\ \citenamefont
  {Abanin}}]{PhysRevLett.110.260601}%
  \BibitemOpen
  \bibfield  {author} {\bibinfo {author} {\bibfnamefont {M.}~\bibnamefont
  {Serbyn}}, \bibinfo {author} {\bibfnamefont {Z.}~\bibnamefont
  {Papi\ifmmode~\acute{c}\else \'{c}\fi{}}}, \ and\ \bibinfo {author}
  {\bibfnamefont {D.~A.}\ \bibnamefont {Abanin}},\ }\href {\doibase
  10.1103/PhysRevLett.110.260601} {\bibfield  {journal} {\bibinfo  {journal}
  {Phys. Rev. Lett.}\ }\textbf {\bibinfo {volume} {110}},\ \bibinfo {pages}
  {260601} (\bibinfo {year} {2013}{\natexlab{a}})}\BibitemShut {NoStop}%
\bibitem [{\citenamefont {Serbyn}\ \emph
  {et~al.}(2014{\natexlab{a}})\citenamefont {Serbyn}, \citenamefont {Knap},
  \citenamefont {Gopalakrishnan}, \citenamefont {Papi\ifmmode~\acute{c}\else
  \'{c}\fi{}}, \citenamefont {Yao}, \citenamefont {Laumann}, \citenamefont
  {Abanin}, \citenamefont {Lukin},\ and\ \citenamefont
  {Demler}}]{PhysRevLett.113.147204}%
  \BibitemOpen
  \bibfield  {author} {\bibinfo {author} {\bibfnamefont {M.}~\bibnamefont
  {Serbyn}}, \bibinfo {author} {\bibfnamefont {M.}~\bibnamefont {Knap}},
  \bibinfo {author} {\bibfnamefont {S.}~\bibnamefont {Gopalakrishnan}},
  \bibinfo {author} {\bibfnamefont {Z.}~\bibnamefont
  {Papi\ifmmode~\acute{c}\else \'{c}\fi{}}}, \bibinfo {author} {\bibfnamefont
  {N.~Y.}\ \bibnamefont {Yao}}, \bibinfo {author} {\bibfnamefont {C.~R.}\
  \bibnamefont {Laumann}}, \bibinfo {author} {\bibfnamefont {D.~A.}\
  \bibnamefont {Abanin}}, \bibinfo {author} {\bibfnamefont {M.~D.}\
  \bibnamefont {Lukin}}, \ and\ \bibinfo {author} {\bibfnamefont {E.~A.}\
  \bibnamefont {Demler}},\ }\href {\doibase 10.1103/PhysRevLett.113.147204}
  {\bibfield  {journal} {\bibinfo  {journal} {Phys. Rev. Lett.}\ }\textbf
  {\bibinfo {volume} {113}},\ \bibinfo {pages} {147204} (\bibinfo {year}
  {2014}{\natexlab{a}})}\BibitemShut {NoStop}%
\bibitem [{\citenamefont {Vasseur}\ \emph
  {et~al.}(2015{\natexlab{a}})\citenamefont {Vasseur}, \citenamefont
  {Parameswaran},\ and\ \citenamefont {Moore}}]{PhysRevB.91.140202}%
  \BibitemOpen
  \bibfield  {author} {\bibinfo {author} {\bibfnamefont {R.}~\bibnamefont
  {Vasseur}}, \bibinfo {author} {\bibfnamefont {S.~A.}\ \bibnamefont
  {Parameswaran}}, \ and\ \bibinfo {author} {\bibfnamefont {J.~E.}\
  \bibnamefont {Moore}},\ }\href {\doibase 10.1103/PhysRevB.91.140202}
  {\bibfield  {journal} {\bibinfo  {journal} {Phys. Rev. B}\ }\textbf {\bibinfo
  {volume} {91}},\ \bibinfo {pages} {140202} (\bibinfo {year}
  {2015}{\natexlab{a}})}\BibitemShut {NoStop}%
\bibitem [{\citenamefont {Bahri}\ \emph {et~al.}(2015)\citenamefont {Bahri},
  \citenamefont {Vosk}, \citenamefont {Altman},\ and\ \citenamefont
  {Vishwanath}}]{BahriMBLSPT}%
  \BibitemOpen
  \bibfield  {author} {\bibinfo {author} {\bibfnamefont {Y.}~\bibnamefont
  {Bahri}}, \bibinfo {author} {\bibfnamefont {R.}~\bibnamefont {Vosk}},
  \bibinfo {author} {\bibfnamefont {E.}~\bibnamefont {Altman}}, \ and\ \bibinfo
  {author} {\bibfnamefont {A.}~\bibnamefont {Vishwanath}},\ }\href
  {http://dx.doi.org/10.1038/ncomms8341} {\bibfield  {journal} {\bibinfo
  {journal} {Nat. Commun.}\ }\textbf {\bibinfo {volume} {6}} (\bibinfo {year}
  {2015})}\BibitemShut {NoStop}%
\bibitem [{\citenamefont {Serbyn}\ \emph
  {et~al.}(2014{\natexlab{b}})\citenamefont {Serbyn}, \citenamefont
  {Papi\ifmmode~\acute{c}\else \'{c}\fi{}},\ and\ \citenamefont
  {Abanin}}]{PhysRevB.90.174302}%
  \BibitemOpen
  \bibfield  {author} {\bibinfo {author} {\bibfnamefont {M.}~\bibnamefont
  {Serbyn}}, \bibinfo {author} {\bibfnamefont {Z.}~\bibnamefont
  {Papi\ifmmode~\acute{c}\else \'{c}\fi{}}}, \ and\ \bibinfo {author}
  {\bibfnamefont {D.~A.}\ \bibnamefont {Abanin}},\ }\href {\doibase
  10.1103/PhysRevB.90.174302} {\bibfield  {journal} {\bibinfo  {journal} {Phys.
  Rev. B}\ }\textbf {\bibinfo {volume} {90}},\ \bibinfo {pages} {174302}
  (\bibinfo {year} {2014}{\natexlab{b}})}\BibitemShut {NoStop}%
\bibitem [{\citenamefont {Huse}\ \emph {et~al.}(2013)\citenamefont {Huse},
  \citenamefont {Nandkishore}, \citenamefont {Oganesyan}, \citenamefont {Pal},\
  and\ \citenamefont {Sondhi}}]{HuseMBLQuantumOrder}%
  \BibitemOpen
  \bibfield  {author} {\bibinfo {author} {\bibfnamefont {D.~A.}\ \bibnamefont
  {Huse}}, \bibinfo {author} {\bibfnamefont {R.}~\bibnamefont {Nandkishore}},
  \bibinfo {author} {\bibfnamefont {V.}~\bibnamefont {Oganesyan}}, \bibinfo
  {author} {\bibfnamefont {A.}~\bibnamefont {Pal}}, \ and\ \bibinfo {author}
  {\bibfnamefont {S.~L.}\ \bibnamefont {Sondhi}},\ }\href {\doibase
  10.1103/PhysRevB.88.014206} {\bibfield  {journal} {\bibinfo  {journal} {Phys.
  Rev. B}\ }\textbf {\bibinfo {volume} {88}},\ \bibinfo {pages} {014206}
  (\bibinfo {year} {2013})}\BibitemShut {NoStop}%
\bibitem [{\citenamefont {Chandran}\ \emph {et~al.}(2014)\citenamefont
  {Chandran}, \citenamefont {Khemani}, \citenamefont {Laumann},\ and\
  \citenamefont {Sondhi}}]{PhysRevB.89.144201}%
  \BibitemOpen
  \bibfield  {author} {\bibinfo {author} {\bibfnamefont {A.}~\bibnamefont
  {Chandran}}, \bibinfo {author} {\bibfnamefont {V.}~\bibnamefont {Khemani}},
  \bibinfo {author} {\bibfnamefont {C.~R.}\ \bibnamefont {Laumann}}, \ and\
  \bibinfo {author} {\bibfnamefont {S.~L.}\ \bibnamefont {Sondhi}},\ }\href
  {\doibase 10.1103/PhysRevB.89.144201} {\bibfield  {journal} {\bibinfo
  {journal} {Phys. Rev. B}\ }\textbf {\bibinfo {volume} {89}},\ \bibinfo
  {pages} {144201} (\bibinfo {year} {2014})}\BibitemShut {NoStop}%
\bibitem [{\citenamefont {Pekker}\ \emph {et~al.}(2014)\citenamefont {Pekker},
  \citenamefont {Refael}, \citenamefont {Altman}, \citenamefont {Demler},\ and\
  \citenamefont {Oganesyan}}]{PekkerRSRGX}%
  \BibitemOpen
  \bibfield  {author} {\bibinfo {author} {\bibfnamefont {D.}~\bibnamefont
  {Pekker}}, \bibinfo {author} {\bibfnamefont {G.}~\bibnamefont {Refael}},
  \bibinfo {author} {\bibfnamefont {E.}~\bibnamefont {Altman}}, \bibinfo
  {author} {\bibfnamefont {E.}~\bibnamefont {Demler}}, \ and\ \bibinfo {author}
  {\bibfnamefont {V.}~\bibnamefont {Oganesyan}},\ }\href {\doibase
  10.1103/PhysRevX.4.011052} {\bibfield  {journal} {\bibinfo  {journal} {Phys.
  Rev. X}\ }\textbf {\bibinfo {volume} {4}},\ \bibinfo {pages} {011052}
  (\bibinfo {year} {2014})}\BibitemShut {NoStop}%
\bibitem [{\citenamefont {Kj\"all}\ \emph {et~al.}(2014)\citenamefont
  {Kj\"all}, \citenamefont {Bardarson},\ and\ \citenamefont
  {Pollmann}}]{PhysRevLett.113.107204}%
  \BibitemOpen
  \bibfield  {author} {\bibinfo {author} {\bibfnamefont {J.~A.}\ \bibnamefont
  {Kj\"all}}, \bibinfo {author} {\bibfnamefont {J.~H.}\ \bibnamefont
  {Bardarson}}, \ and\ \bibinfo {author} {\bibfnamefont {F.}~\bibnamefont
  {Pollmann}},\ }\href {\doibase 10.1103/PhysRevLett.113.107204} {\bibfield
  {journal} {\bibinfo  {journal} {Phys. Rev. Lett.}\ }\textbf {\bibinfo
  {volume} {113}},\ \bibinfo {pages} {107204} (\bibinfo {year}
  {2014})}\BibitemShut {NoStop}%
\bibitem [{\citenamefont {Luitz}\ \emph {et~al.}(2015)\citenamefont {Luitz},
  \citenamefont {Laflorencie},\ and\ \citenamefont {Alet}}]{Luitz}%
  \BibitemOpen
  \bibfield  {author} {\bibinfo {author} {\bibfnamefont {D.~J.}\ \bibnamefont
  {Luitz}}, \bibinfo {author} {\bibfnamefont {N.}~\bibnamefont {Laflorencie}},
  \ and\ \bibinfo {author} {\bibfnamefont {F.}~\bibnamefont {Alet}},\ }\href
  {\doibase 10.1103/PhysRevB.91.081103} {\bibfield  {journal} {\bibinfo
  {journal} {Phys. Rev. B}\ }\textbf {\bibinfo {volume} {91}},\ \bibinfo
  {pages} {081103} (\bibinfo {year} {2015})}\BibitemShut {NoStop}%
\bibitem [{\citenamefont {Luitz}\ \emph {et~al.}(2016)\citenamefont {Luitz},
  \citenamefont {Laflorencie},\ and\ \citenamefont
  {Alet}}]{PhysRevB.93.060201}%
  \BibitemOpen
  \bibfield  {author} {\bibinfo {author} {\bibfnamefont {D.~J.}\ \bibnamefont
  {Luitz}}, \bibinfo {author} {\bibfnamefont {N.}~\bibnamefont {Laflorencie}},
  \ and\ \bibinfo {author} {\bibfnamefont {F.}~\bibnamefont {Alet}},\ }\href
  {\doibase 10.1103/PhysRevB.93.060201} {\bibfield  {journal} {\bibinfo
  {journal} {Phys. Rev. B}\ }\textbf {\bibinfo {volume} {93}},\ \bibinfo
  {pages} {060201} (\bibinfo {year} {2016})}\BibitemShut {NoStop}%
\bibitem [{\citenamefont {Khemani}\ \emph {et~al.}(2017)\citenamefont
  {Khemani}, \citenamefont {Lim}, \citenamefont {Sheng},\ and\ \citenamefont
  {Huse}}]{khemani2016critical}%
  \BibitemOpen
  \bibfield  {author} {\bibinfo {author} {\bibfnamefont {V.}~\bibnamefont
  {Khemani}}, \bibinfo {author} {\bibfnamefont {S.~P.}\ \bibnamefont {Lim}},
  \bibinfo {author} {\bibfnamefont {D.~N.}\ \bibnamefont {Sheng}}, \ and\
  \bibinfo {author} {\bibfnamefont {D.~A.}\ \bibnamefont {Huse}},\ }\href
  {\doibase 10.1103/PhysRevX.7.021013} {\bibfield  {journal} {\bibinfo
  {journal} {Phys. Rev. X}\ }\textbf {\bibinfo {volume} {7}},\ \bibinfo {pages}
  {021013} (\bibinfo {year} {2017})}\BibitemShut {NoStop}%
\bibitem [{\citenamefont {Chayes}\ \emph {et~al.}(1986)\citenamefont {Chayes},
  \citenamefont {Chayes}, \citenamefont {Fisher},\ and\ \citenamefont
  {Spencer}}]{Chayes}%
  \BibitemOpen
  \bibfield  {author} {\bibinfo {author} {\bibfnamefont {J.~T.}\ \bibnamefont
  {Chayes}}, \bibinfo {author} {\bibfnamefont {L.}~\bibnamefont {Chayes}},
  \bibinfo {author} {\bibfnamefont {D.~S.}\ \bibnamefont {Fisher}}, \ and\
  \bibinfo {author} {\bibfnamefont {T.}~\bibnamefont {Spencer}},\ }\href
  {\doibase 10.1103/PhysRevLett.57.2999} {\bibfield  {journal} {\bibinfo
  {journal} {Phys. Rev. Lett.}\ }\textbf {\bibinfo {volume} {57}},\ \bibinfo
  {pages} {2999} (\bibinfo {year} {1986})}\BibitemShut {NoStop}%
\bibitem [{\citenamefont {Chandran}\ \emph {et~al.}(2015)\citenamefont
  {Chandran}, \citenamefont {Laumann},\ and\ \citenamefont
  {Oganesyan}}]{chandran2015finite}%
  \BibitemOpen
  \bibfield  {author} {\bibinfo {author} {\bibfnamefont {A.}~\bibnamefont
  {Chandran}}, \bibinfo {author} {\bibfnamefont {C.}~\bibnamefont {Laumann}}, \
  and\ \bibinfo {author} {\bibfnamefont {V.}~\bibnamefont {Oganesyan}},\
  }\href@noop {} {\bibfield  {journal} {\bibinfo  {journal} {arXiv preprint
  arXiv:1509.04285}\ } (\bibinfo {year} {2015})}\BibitemShut {NoStop}%
\bibitem [{\citenamefont {Vosk}\ \emph {et~al.}(2015)\citenamefont {Vosk},
  \citenamefont {Huse},\ and\ \citenamefont {Altman}}]{Vosk:2015aa}%
  \BibitemOpen
  \bibfield  {author} {\bibinfo {author} {\bibfnamefont {R.}~\bibnamefont
  {Vosk}}, \bibinfo {author} {\bibfnamefont {D.~A.}\ \bibnamefont {Huse}}, \
  and\ \bibinfo {author} {\bibfnamefont {E.}~\bibnamefont {Altman}},\ }\href
  {\doibase 10.1103/PhysRevX.5.031032} {\bibfield  {journal} {\bibinfo
  {journal} {Phys. Rev. X}\ }\textbf {\bibinfo {volume} {5}},\ \bibinfo {pages}
  {031032} (\bibinfo {year} {2015})}\BibitemShut {NoStop}%
\bibitem [{\citenamefont {Potter}\ \emph {et~al.}(2015)\citenamefont {Potter},
  \citenamefont {Vasseur},\ and\ \citenamefont {Parameswaran}}]{Potter:2015aa}%
  \BibitemOpen
  \bibfield  {author} {\bibinfo {author} {\bibfnamefont {A.~C.}\ \bibnamefont
  {Potter}}, \bibinfo {author} {\bibfnamefont {R.}~\bibnamefont {Vasseur}}, \
  and\ \bibinfo {author} {\bibfnamefont {S.~A.}\ \bibnamefont {Parameswaran}},\
  }\href {\doibase 10.1103/PhysRevX.5.031033} {\bibfield  {journal} {\bibinfo
  {journal} {Phys. Rev. X}\ }\textbf {\bibinfo {volume} {5}},\ \bibinfo {pages}
  {031033} (\bibinfo {year} {2015})}\BibitemShut {NoStop}%
\bibitem [{\citenamefont {Zhang}\ \emph {et~al.}(2016)\citenamefont {Zhang},
  \citenamefont {Zhao}, \citenamefont {Devakul},\ and\ \citenamefont
  {Huse}}]{PhysRevB.93.224201}%
  \BibitemOpen
  \bibfield  {author} {\bibinfo {author} {\bibfnamefont {L.}~\bibnamefont
  {Zhang}}, \bibinfo {author} {\bibfnamefont {B.}~\bibnamefont {Zhao}},
  \bibinfo {author} {\bibfnamefont {T.}~\bibnamefont {Devakul}}, \ and\
  \bibinfo {author} {\bibfnamefont {D.~A.}\ \bibnamefont {Huse}},\ }\href
  {\doibase 10.1103/PhysRevB.93.224201} {\bibfield  {journal} {\bibinfo
  {journal} {Phys. Rev. B}\ }\textbf {\bibinfo {volume} {93}},\ \bibinfo
  {pages} {224201} (\bibinfo {year} {2016})}\BibitemShut {NoStop}%
\bibitem [{\citenamefont {Serbyn}\ \emph
  {et~al.}(2013{\natexlab{b}})\citenamefont {Serbyn}, \citenamefont
  {Papi\ifmmode~\acute{c}\else \'{c}\fi{}},\ and\ \citenamefont
  {Abanin}}]{PhysRevLett.111.127201}%
  \BibitemOpen
  \bibfield  {author} {\bibinfo {author} {\bibfnamefont {M.}~\bibnamefont
  {Serbyn}}, \bibinfo {author} {\bibfnamefont {Z.}~\bibnamefont
  {Papi\ifmmode~\acute{c}\else \'{c}\fi{}}}, \ and\ \bibinfo {author}
  {\bibfnamefont {D.~A.}\ \bibnamefont {Abanin}},\ }\href {\doibase
  10.1103/PhysRevLett.111.127201} {\bibfield  {journal} {\bibinfo  {journal}
  {Phys. Rev. Lett.}\ }\textbf {\bibinfo {volume} {111}},\ \bibinfo {pages}
  {127201} (\bibinfo {year} {2013}{\natexlab{b}})}\BibitemShut {NoStop}%
\bibitem [{\citenamefont {Huse}\ \emph {et~al.}(2014)\citenamefont {Huse},
  \citenamefont {Nandkishore},\ and\ \citenamefont
  {Oganesyan}}]{PhysRevB.90.174202}%
  \BibitemOpen
  \bibfield  {author} {\bibinfo {author} {\bibfnamefont {D.~A.}\ \bibnamefont
  {Huse}}, \bibinfo {author} {\bibfnamefont {R.}~\bibnamefont {Nandkishore}}, \
  and\ \bibinfo {author} {\bibfnamefont {V.}~\bibnamefont {Oganesyan}},\ }\href
  {\doibase 10.1103/PhysRevB.90.174202} {\bibfield  {journal} {\bibinfo
  {journal} {Phys. Rev. B}\ }\textbf {\bibinfo {volume} {90}},\ \bibinfo
  {pages} {174202} (\bibinfo {year} {2014})}\BibitemShut {NoStop}%
\bibitem [{\citenamefont {Ros}\ \emph {et~al.}(2015)\citenamefont {Ros},
  \citenamefont {M{\"u}ller},\ and\ \citenamefont {Scardicchio}}]{Ros2015420}%
  \BibitemOpen
  \bibfield  {author} {\bibinfo {author} {\bibfnamefont {V.}~\bibnamefont
  {Ros}}, \bibinfo {author} {\bibfnamefont {M.}~\bibnamefont {M{\"u}ller}}, \
  and\ \bibinfo {author} {\bibfnamefont {A.}~\bibnamefont {Scardicchio}},\
  }\href {\doibase http://dx.doi.org/10.1016/j.nuclphysb.2014.12.014}
  {\bibfield  {journal} {\bibinfo  {journal} {Nucl. Phys. B}\ }\textbf
  {\bibinfo {volume} {891}},\ \bibinfo {pages} {420 } (\bibinfo {year}
  {2015})}\BibitemShut {NoStop}%
\bibitem [{\citenamefont {Imbrie}(2016)}]{PhysRevLett.117.027201}%
  \BibitemOpen
  \bibfield  {author} {\bibinfo {author} {\bibfnamefont {J.~Z.}\ \bibnamefont
  {Imbrie}},\ }\href {\doibase 10.1103/PhysRevLett.117.027201} {\bibfield
  {journal} {\bibinfo  {journal} {Phys. Rev. Lett.}\ }\textbf {\bibinfo
  {volume} {117}},\ \bibinfo {pages} {027201} (\bibinfo {year}
  {2016})}\BibitemShut {NoStop}%
\bibitem [{Note1()}]{Note1}%
  \BibitemOpen
  \bibinfo {note} {Unless the level spacing of one cluster exceeds the
  bandwidth of the other; then $\Lambda _{i'} \geq \delta _{i'} \geq \Lambda
  _{j'} \geq \delta _{j'}$ and $\Delta E_{i'j'} = \protect \qopname \relax
  m{max}\left (\delta _{i'} - \Lambda _{j'}, \delta _{j'}\right )$}\BibitemShut
  {NoStop}%
\bibitem [{Sup()}]{SupMat}%
  \BibitemOpen
  \href@noop {} {\bibinfo  {journal} {See supplemental material}\ }\BibitemShut
  {NoStop}%
\bibitem [{\citenamefont {De~Roeck}\ and\ \citenamefont
  {Huveneers}(2017)}]{2016arXiv160801815D}%
  \BibitemOpen
\bibfield  {journal} {  }\bibfield  {author} {\bibinfo {author} {\bibfnamefont
  {W.}~\bibnamefont {De~Roeck}}\ and\ \bibinfo {author} {\bibfnamefont
  {F.}~\bibnamefont {Huveneers}},\ }\href {\doibase 10.1103/PhysRevB.95.155129}
  {\bibfield  {journal} {\bibinfo  {journal} {Phys. Rev. B}\ }\textbf {\bibinfo
  {volume} {95}},\ \bibinfo {pages} {155129} (\bibinfo {year}
  {2017})}\BibitemShut {NoStop}%
\bibitem [{\citenamefont {Serbyn}\ \emph {et~al.}(2015)\citenamefont {Serbyn},
  \citenamefont {Papi\ifmmode~\acute{c}\else \'{c}\fi{}},\ and\ \citenamefont
  {Abanin}}]{PhysRevX.5.041047}%
  \BibitemOpen
  \bibfield  {author} {\bibinfo {author} {\bibfnamefont {M.}~\bibnamefont
  {Serbyn}}, \bibinfo {author} {\bibfnamefont {Z.}~\bibnamefont
  {Papi\ifmmode~\acute{c}\else \'{c}\fi{}}}, \ and\ \bibinfo {author}
  {\bibfnamefont {D.~A.}\ \bibnamefont {Abanin}},\ }\href {\doibase
  10.1103/PhysRevX.5.041047} {\bibfield  {journal} {\bibinfo  {journal} {Phys.
  Rev. X}\ }\textbf {\bibinfo {volume} {5}},\ \bibinfo {pages} {041047}
  (\bibinfo {year} {2015})}\BibitemShut {NoStop}%
\bibitem [{\citenamefont {Fisher}(1992)}]{FisherRSRG1}%
  \BibitemOpen
  \bibfield  {author} {\bibinfo {author} {\bibfnamefont {D.~S.}\ \bibnamefont
  {Fisher}},\ }\href {\doibase 10.1103/PhysRevLett.69.534} {\bibfield
  {journal} {\bibinfo  {journal} {Phys. Rev. Lett.}\ }\textbf {\bibinfo
  {volume} {69}},\ \bibinfo {pages} {534} (\bibinfo {year} {1992})}\BibitemShut
  {NoStop}%
\bibitem [{\citenamefont {Fisher}(1994)}]{FisherRSRG2}%
  \BibitemOpen
  \bibfield  {author} {\bibinfo {author} {\bibfnamefont {D.~S.}\ \bibnamefont
  {Fisher}},\ }\href {\doibase 10.1103/PhysRevB.50.3799} {\bibfield  {journal}
  {\bibinfo  {journal} {Phys. Rev. B}\ }\textbf {\bibinfo {volume} {50}},\
  \bibinfo {pages} {3799} (\bibinfo {year} {1994})}\BibitemShut {NoStop}%
\bibitem [{\citenamefont {Vosk}\ and\ \citenamefont
  {Altman}(2014)}]{PhysRevLett.112.217204}%
  \BibitemOpen
  \bibfield  {author} {\bibinfo {author} {\bibfnamefont {R.}~\bibnamefont
  {Vosk}}\ and\ \bibinfo {author} {\bibfnamefont {E.}~\bibnamefont {Altman}},\
  }\href {\doibase 10.1103/PhysRevLett.112.217204} {\bibfield  {journal}
  {\bibinfo  {journal} {Phys. Rev. Lett.}\ }\textbf {\bibinfo {volume} {112}},\
  \bibinfo {pages} {217204} (\bibinfo {year} {2014})}\BibitemShut {NoStop}%
\bibitem [{\citenamefont {Vasseur}\ \emph
  {et~al.}(2015{\natexlab{b}})\citenamefont {Vasseur}, \citenamefont {Potter},\
  and\ \citenamefont {Parameswaran}}]{QCGPRL}%
  \BibitemOpen
  \bibfield  {author} {\bibinfo {author} {\bibfnamefont {R.}~\bibnamefont
  {Vasseur}}, \bibinfo {author} {\bibfnamefont {A.~C.}\ \bibnamefont {Potter}},
  \ and\ \bibinfo {author} {\bibfnamefont {S.~A.}\ \bibnamefont
  {Parameswaran}},\ }\href {\doibase 10.1103/PhysRevLett.114.217201} {\bibfield
   {journal} {\bibinfo  {journal} {Phys. Rev. Lett.}\ }\textbf {\bibinfo
  {volume} {114}},\ \bibinfo {pages} {217201} (\bibinfo {year}
  {2015}{\natexlab{b}})}\BibitemShut {NoStop}%
\bibitem [{\citenamefont {Agarwal}\ \emph {et~al.}(2017)\citenamefont
  {Agarwal}, \citenamefont {Altman}, \citenamefont {Demler}, \citenamefont
  {Gopalakrishnan}, \citenamefont {Huse},\ and\ \citenamefont
  {Knap}}]{2016arXiv161100770A}%
  \BibitemOpen
  \bibfield  {author} {\bibinfo {author} {\bibfnamefont {K.}~\bibnamefont
  {Agarwal}}, \bibinfo {author} {\bibfnamefont {E.}~\bibnamefont {Altman}},
  \bibinfo {author} {\bibfnamefont {E.}~\bibnamefont {Demler}}, \bibinfo
  {author} {\bibfnamefont {S.}~\bibnamefont {Gopalakrishnan}}, \bibinfo
  {author} {\bibfnamefont {D.~A.}\ \bibnamefont {Huse}}, \ and\ \bibinfo
  {author} {\bibfnamefont {M.}~\bibnamefont {Knap}},\ }\href {\doibase
  10.1002/andp.201600326} {\bibfield  {journal} {\bibinfo  {journal} {Ann.
  Phys. (Berlin)}\ }\textbf {\bibinfo {volume} {529}},\ \bibinfo {pages}
  {1600326} (\bibinfo {year} {2017})}\BibitemShut {NoStop}%
\bibitem [{Note2()}]{Note2}%
  \BibitemOpen
  \bibinfo {note} {Re-examining~\cite {Potter:2015aa} showed that those RG
  rules allowed rare avalanches; this does not quantitively alter the reported
  critical properties.}\BibitemShut {Stop}%
\bibitem [{\citenamefont {Yu}\ \emph {et~al.}(2016)\citenamefont {Yu},
  \citenamefont {Luitz},\ and\ \citenamefont {Clark}}]{yu2016bimodal}%
  \BibitemOpen
  \bibfield  {author} {\bibinfo {author} {\bibfnamefont {X.}~\bibnamefont
  {Yu}}, \bibinfo {author} {\bibfnamefont {D.~J.}\ \bibnamefont {Luitz}}, \
  and\ \bibinfo {author} {\bibfnamefont {B.~K.}\ \bibnamefont {Clark}},\ }\href
  {\doibase 10.1103/PhysRevB.94.184202} {\bibfield  {journal} {\bibinfo
  {journal} {Phys. Rev. B}\ }\textbf {\bibinfo {volume} {94}},\ \bibinfo
  {pages} {184202} (\bibinfo {year} {2016})}\BibitemShut {NoStop}%
\bibitem [{\citenamefont {{Grover}}(2014)}]{2014arXiv1405.1471G}%
  \BibitemOpen
  \bibfield  {author} {\bibinfo {author} {\bibfnamefont {T.}~\bibnamefont
  {{Grover}}},\ }\href@noop {} {\bibfield  {journal} {\bibinfo  {journal}
  {ArXiv e-prints}\ } (\bibinfo {year} {2014})},\ \Eprint
  {http://arxiv.org/abs/1405.1471} {arXiv:1405.1471 [cond-mat.dis-nn]}
  \BibitemShut {NoStop}%
\bibitem [{\citenamefont {Myers}\ and\ \citenamefont
  {Singh}(2012)}]{Myers2012}%
  \BibitemOpen
  \bibfield  {author} {\bibinfo {author} {\bibfnamefont {R.~C.}\ \bibnamefont
  {Myers}}\ and\ \bibinfo {author} {\bibfnamefont {A.}~\bibnamefont {Singh}},\
  }\href {\doibase 10.1007/JHEP04(2012)122} {\bibfield  {journal} {\bibinfo
  {journal} {J. High Energy Phys.}\ }\textbf {\bibinfo {volume} {2012}},\
  \bibinfo {pages} {122} (\bibinfo {year} {2012})}\BibitemShut {NoStop}%
\bibitem [{\citenamefont {Liu}\ and\ \citenamefont {Mezei}(2013)}]{Liu2013}%
  \BibitemOpen
  \bibfield  {author} {\bibinfo {author} {\bibfnamefont {H.}~\bibnamefont
  {Liu}}\ and\ \bibinfo {author} {\bibfnamefont {M.}~\bibnamefont {Mezei}},\
  }\href {\doibase 10.1007/JHEP04(2013)162} {\bibfield  {journal} {\bibinfo
  {journal} {J. High Energy Phys.}\ }\textbf {\bibinfo {volume} {2013}},\
  \bibinfo {pages} {162} (\bibinfo {year} {2013})}\BibitemShut {NoStop}%
\bibitem [{\citenamefont {Huse}(2016)}]{HusePC}%
  \BibitemOpen
  \bibfield  {author} {\bibinfo {author} {\bibfnamefont {D.~A.}\ \bibnamefont
  {Huse}},\ }\href@noop {} {\bibfield  {journal} {\bibinfo  {journal} {Private
  Communication}\ } (\bibinfo {year} {2016})}\BibitemShut {NoStop}%
\end{thebibliography}%

\appendix
\onecolumngrid
\newpage
\section*{Supplementary material}

\twocolumngrid

\subsection{Matrix Element Renormalization}

The rules for renormalizing matrix elements are different from those previously used by \cite{Vosk:2015aa,Potter:2015aa}. 
We discuss additional details which motivate these changes.

We note \cite{PhysRevE.50.888} that the off-diagonal matrix element of a local operator $O$, satisfying ETH is $O_{ij} = \eta_{ij} \sqrt{o(E)/\mathcal{H}}$; $o$ depends on energy, $\eta_{ij}$ are zero-mean univariate random variables, and $\mathcal{H}$ is the number of accessible states.

In the RG, the state of the system is defined by clusters -- resonances between local degrees of freedom. The clusters either grow to describe large thermal puddles or remain small to describe the decoupled local integrals of motion (LIOMs) which characterize the MBL phase.

\begin{figure}[h]
\centering
\subfigure[]{\includegraphics[width=0.4\textwidth]{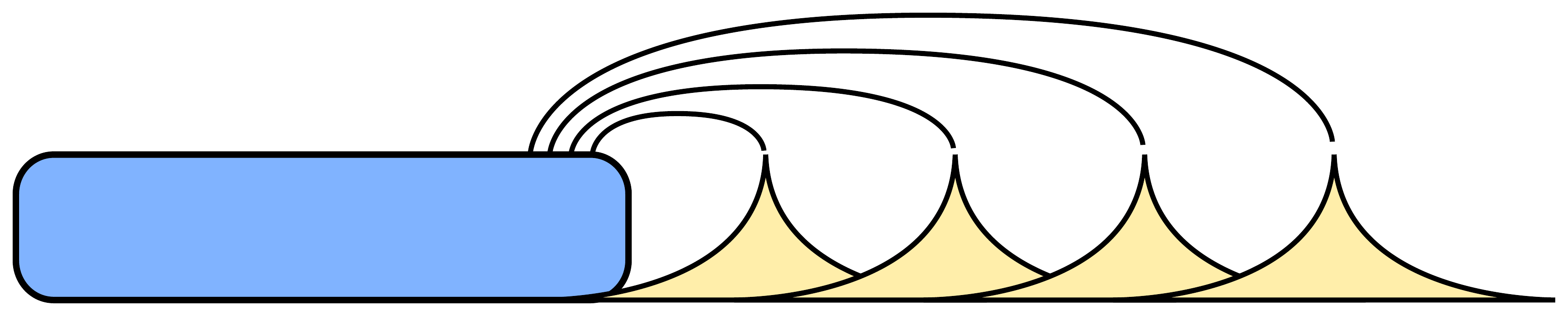}}
\qquad
\subfigure[]{\includegraphics[width=0.4\textwidth]{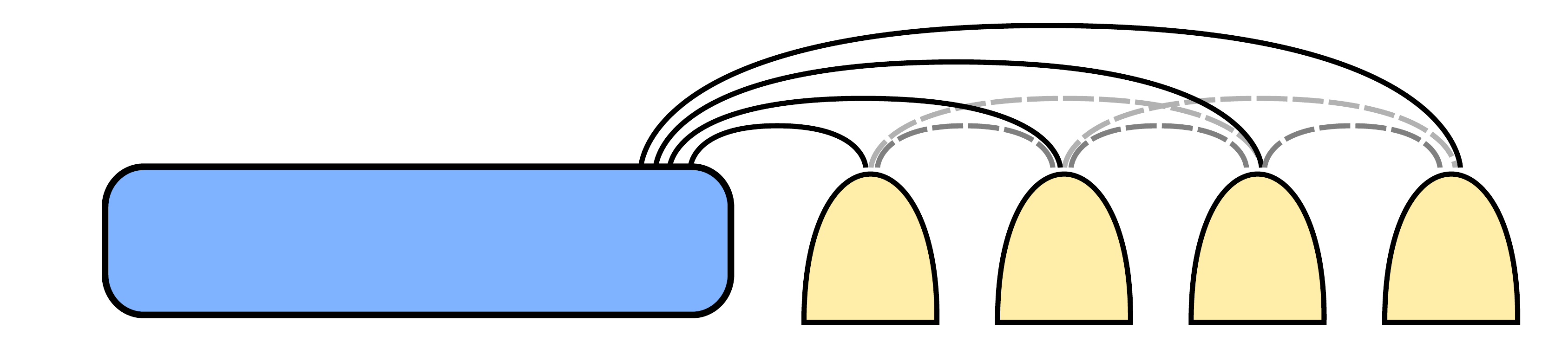}}
\caption{{\bf Schematic of coupling to a small bath. -- } \label{fig:bathcoupling} 
(a) A thermal cluster of inter-resonating sites (blue) interacting with MBL LIOMs (yellow). 
(b) A growing, thermal cluster (blue) interacting with other clusters (yellow) at some stage in the RG procedure. In the RG procedure, clusters change to become either more thermal or more LIOM like.
}
\end{figure}

To understand the construction of resonances, consider the thought experiment of bringing a thermal bath into contact with a MBL system; see Fig.~\ref{fig:bathcoupling}(a). Here the bath can be thought of as a cluster with matrix elements and level statistics well-described by the ETH forms. Before coupling to the bath, the MBL system is composed of LIOMs that interact only classically and cannot change each other's internal state (resonate). The bath couples only locally to the edge of the MBL chain at position $x=0$. Therefore, the overlap amplitude with a LIOM localized near position $x_i$ is $\sim \exp\({-\vert x_i \vert/x_0}\)$, where $x_0$ is the localization length. The bath can now mediate resonances between a pair of LIOMs $i,j$, but these occur only on time scales $\gtrsim \exp\[{(\vert x_i \vert+\vert x_j \vert)/x_0}\]$. In particular, for LIOMs far from the edge of the bath, the time scale will be exponentially long, no matter how close the LIOMs are to each other. It is important to note that the resonance timescale is not related to the coupling of the microscopic particles at sites $i,j$. This scenario was discussed by \cite{HusePC, 2016arXiv160801815D}, and is borne out by numerical simulations \cite{2016arXiv160801815D}.

Consider how these features may be incorporated into the RG. The coupling $\Gamma_{ij}$ is effectively the inverse time scale on which clusters $i,j$ can resonate. As the RG progresses, we consider resonances happening on ever longer time scales. Early in the RG, resonances are fast and occur between a few, strongly-coupled degrees of freedom. Later in the RG, the $\Gamma_{ij}$ have decreased and resonances now occur over longer time scales and involve more degrees of freedom.
This makes precise the sense in which clusters become more `LIOM-like' or more `thermal-like': for each RG step, we determine which clusters can resonate during the corresponding time interval. A pair of clusters may not be able to resonate during that interval and behave relatively `LIOM-like'. In the absence of other clusters, they remain non-resonant, but as part of a larger system they might still be able to resonate at later times if mediated by other clusters.

Let us construct the RG rules for merging clusters to parallel the thought experiment of bona fide LIOMs discussed above. Figure~\ref{fig:bathcoupling}(b) depicts four RG clusters (yellow) which have small $\Gamma_{ij}$ and could not resonate with any other cluster at the pictured RG step. The yellow clusters are also coupled to a cluster (blue) which grew larger and more thermal during the RG step. The off-resonant $\Gamma_{ij}$ between yellow clusters cannot mediate any further direct resonances between these clusters, just as the local microscopic particle coupling cannot drive resonances between LIOMs. While these couplings would, in a more detailed full solution, cause some additional virtual dressing to the cluster, this does not affect the RG or universal properties; we set these couplings $\Gamma_{ij}\rightarrow 0$. In contrast, the coupling to the larger thermal cluster (blue) should be retained. If the blue cluster grows sufficiently thermal, it may, at a later time, thermalize the yellow clusters as part of a large collective resonance. Whether this occurs will be determined later in the RG.

Since the blue cluster grew during the RG step, we must renormalize the coupling between it and the yellow clusters. Indeed, any yellow cluster was coupled to the multiple smaller clusters that merged to form the blue cluster. The ETH assumptions \cite{PhysRevE.50.888} and recent numerical simulations \cite{2016arXiv160801815D} indicate that for the purposes of level spacing and statistics, the blue cluster acts like a fully thermal system spread over all constituent sites that were merged into it. However, the matrix element between the yellow and blue cluster does depend on the original structure of the blue cluster constituents. In particular, the new blue-yellow cluster coupling is dominated by the maximal coupling connecting the original clusters. If we parameterize the final blue-yellow coupling as $\sim \exp(-x_\mathrm{eff}/x_0)$, the effective tunneling distance  $x_\mathrm{eff}$ can be much larger than the geometric separation between the yellow cluster and the edge of the blue bath; see e.g.~\cite{2016arXiv160801815D}. Together the level spacing and matrix element contribution give rise to the renormalization rule \eqref{eq:matRGrule}.

\newpage

\begin{figure*}[h]
\centering
\includegraphics[width=0.84\columnwidth]{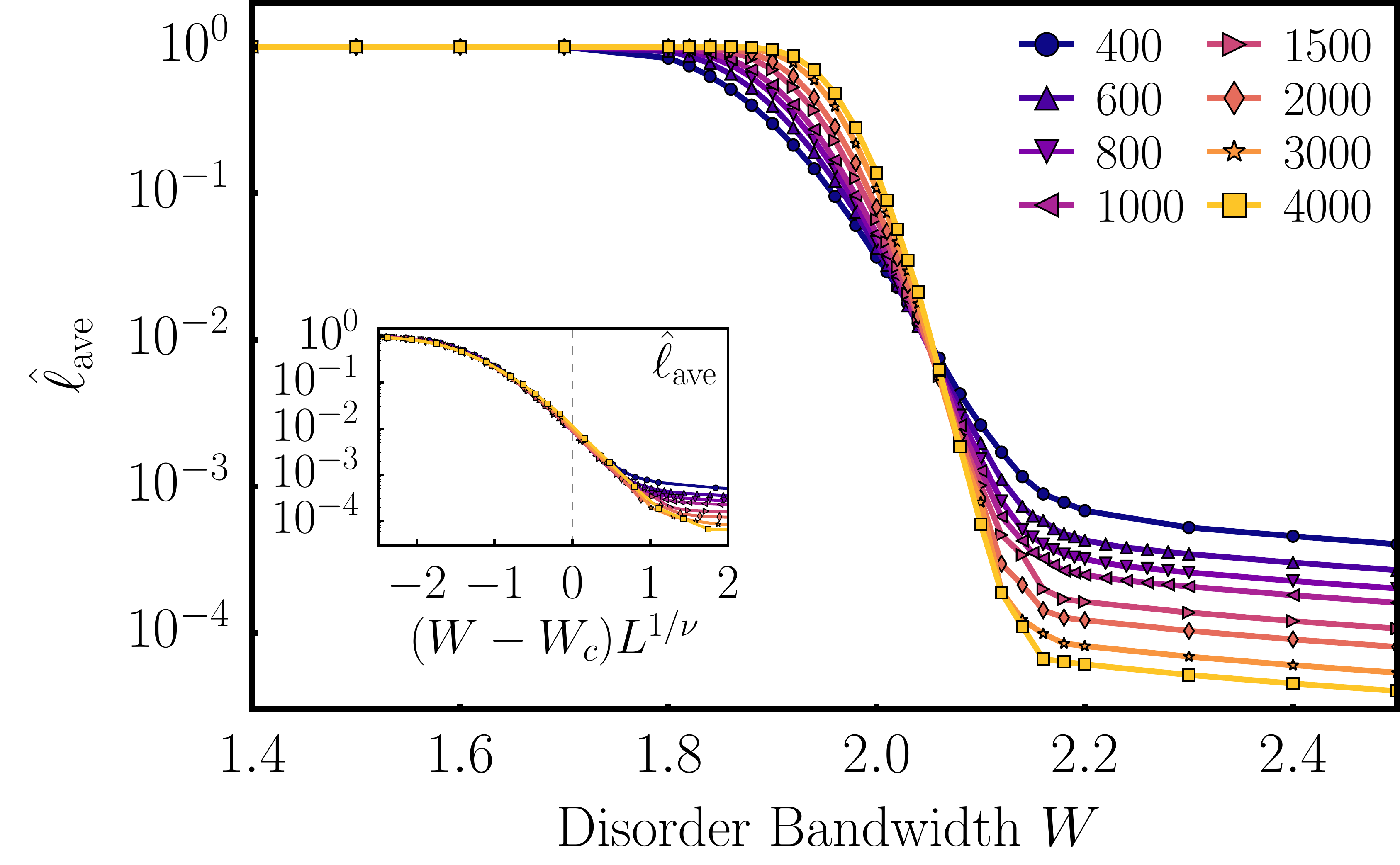}
\includegraphics[width=0.84\columnwidth]{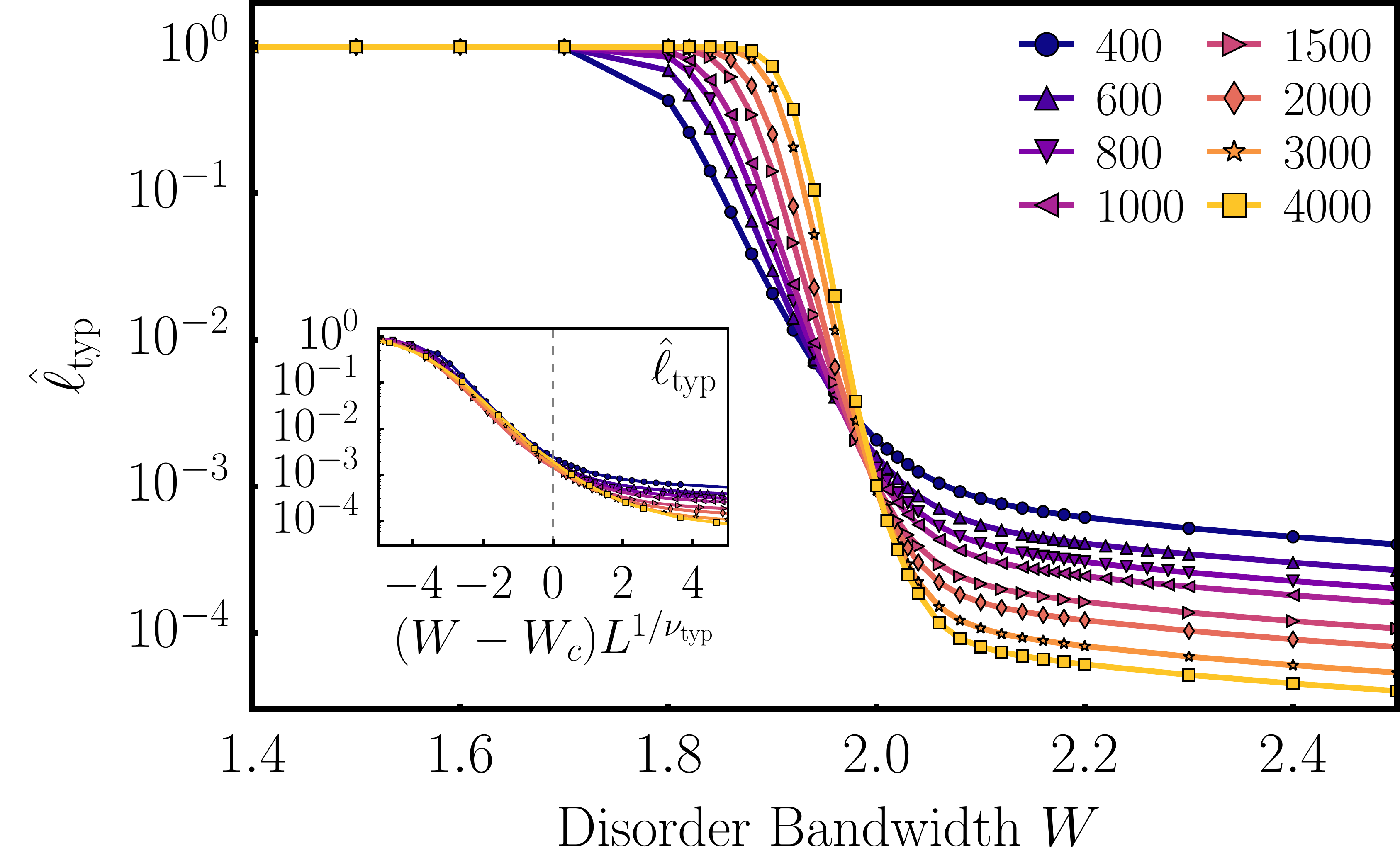}
\caption{ 
Normalized average $\hat{\ell}_{\mathrm{ave}}$ and typical $\hat{\ell}_{\mathrm{typ}}$ cluster length. The typical length ${\ell}_{\mathrm{typ}} = \exp \overline{\[\log\ell\]} $, where $\overline{[\ldots]}$ indicates disorder averaging. The normalization $\hat{\ell}_{\mathrm{x}} = (\ell_{\mathrm{x}} - 1 ) / (L - 1)$ subtracts the contribution of single site clusters. Insets:~scaling collapses with $\nu = 3.2$ and $W_c = 2.05$ [ave] and $\nu_{\mathrm{typ}} = 2.1$ and $W_{c} = 1.99$ [typ]. The typical plot has large finite size corrections and the crossing points drift to higher $W$; the  typical and average crossings should agree with the true transition as $L\to\infty$. Nonetheless, the power $\nu_{\mathrm{typ}} = 2.1 \pm 0.3$ is consistent with extrapolations along cuts of constant $\hat{\ell}$ away from the crossing point itself. The difference in average and typical scaling is characteristic of a broad distribution and indicates that the transition is rare events driven; this is necessary at an infinite randomness transition.
}
\end{figure*}

\begin{figure*}[ht]
\centering
\includegraphics[width=0.82\columnwidth]{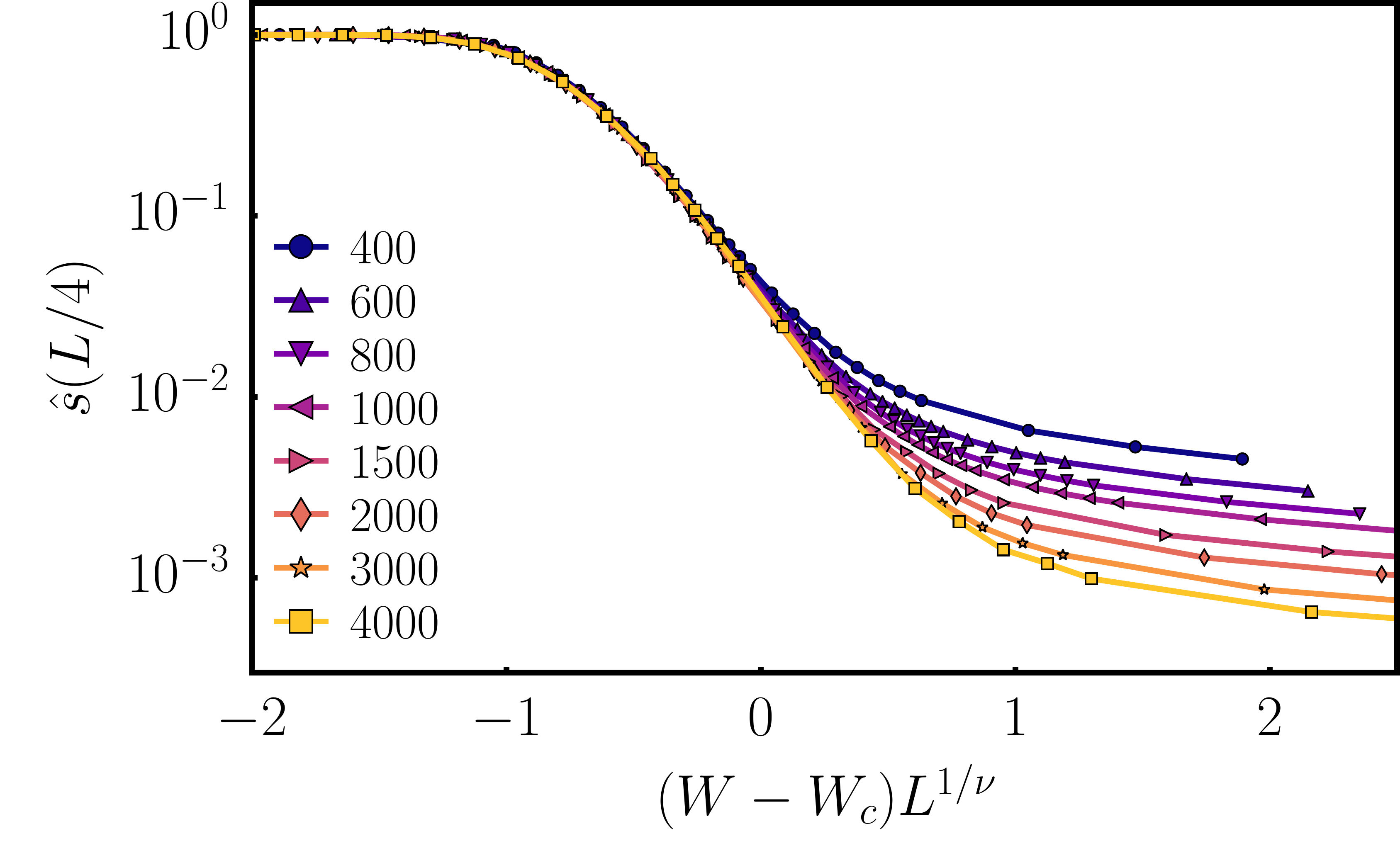}
\includegraphics[width=0.82\columnwidth]{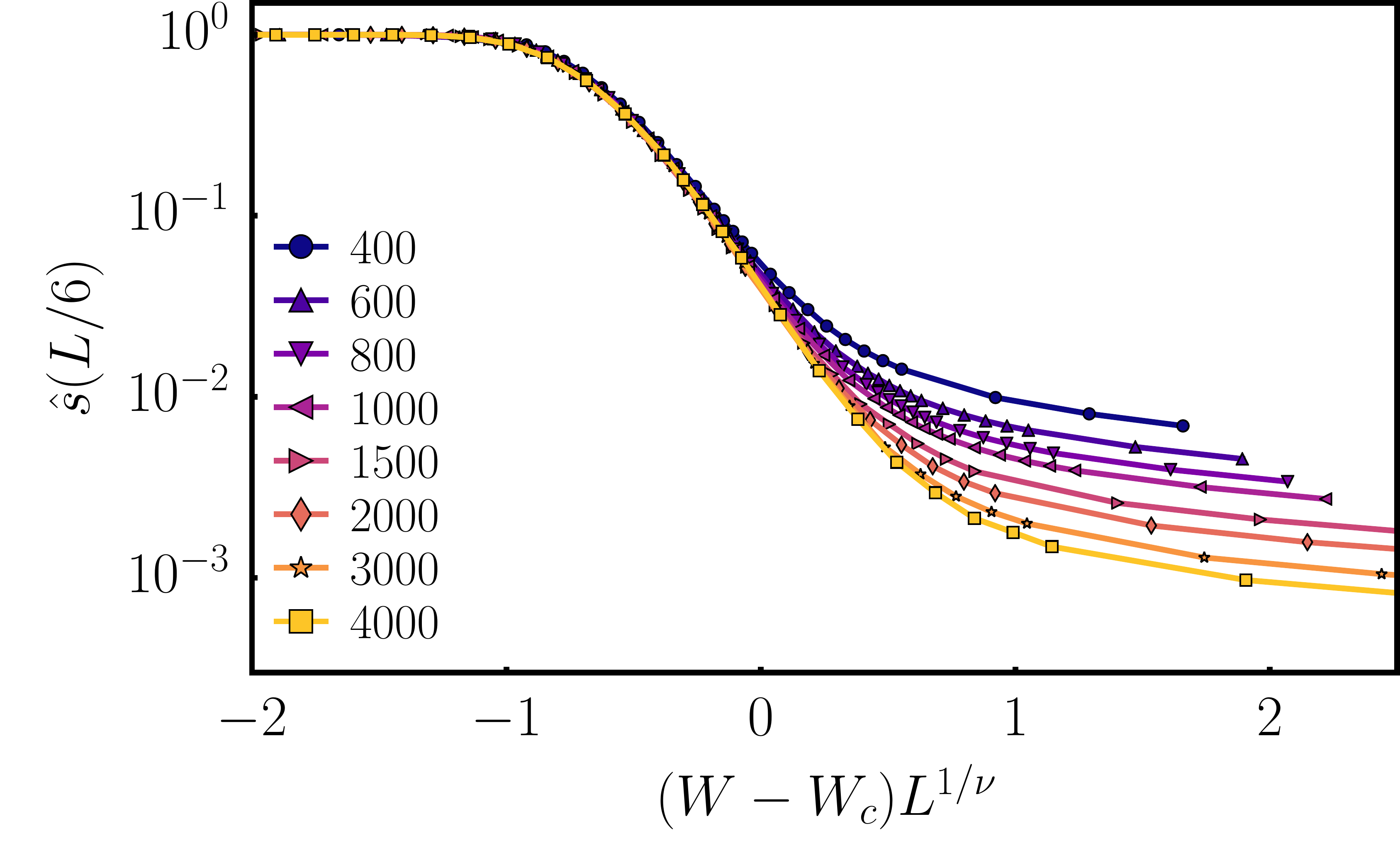}
\includegraphics[width=0.82\columnwidth]{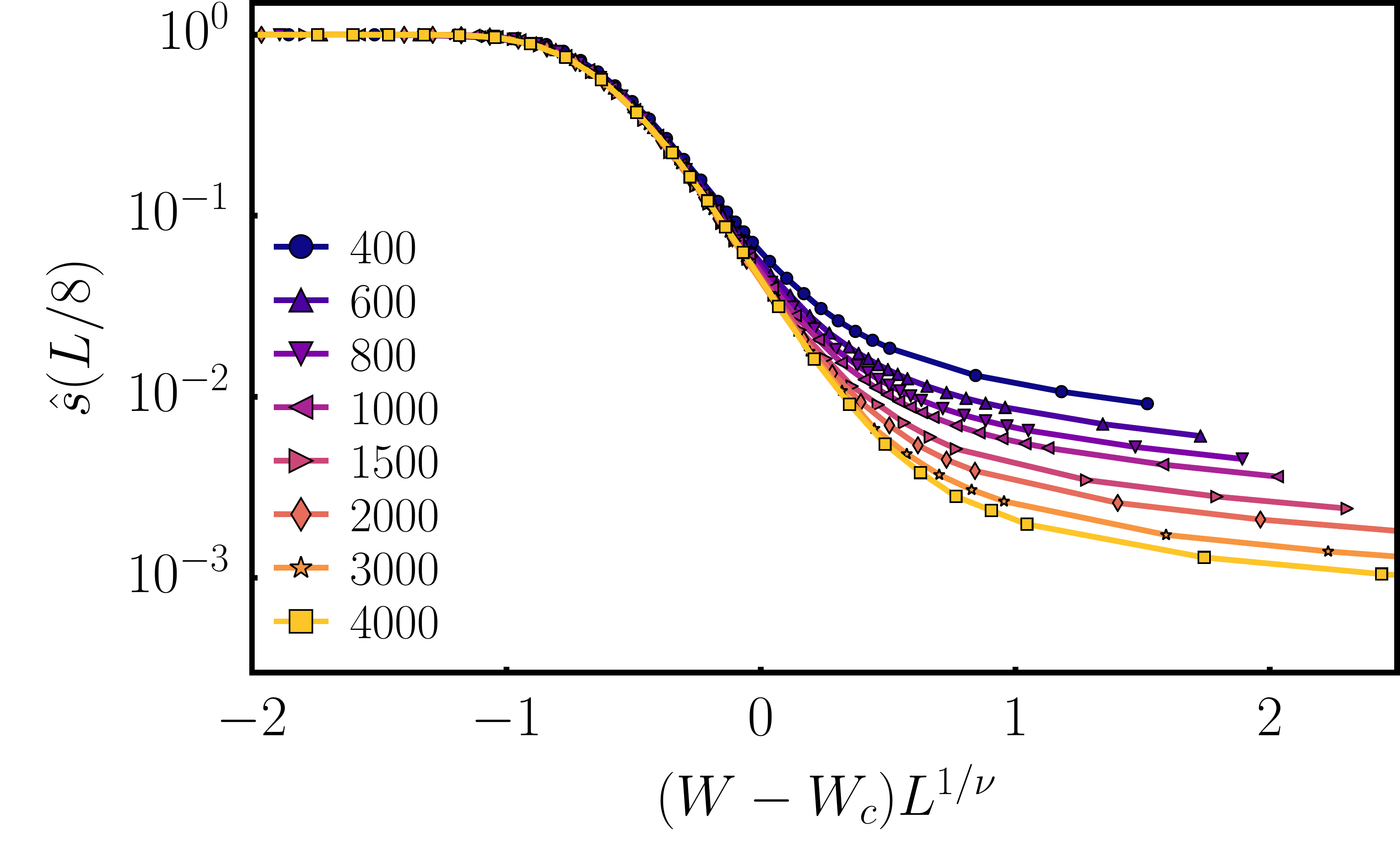}
\includegraphics[width=0.82\columnwidth]{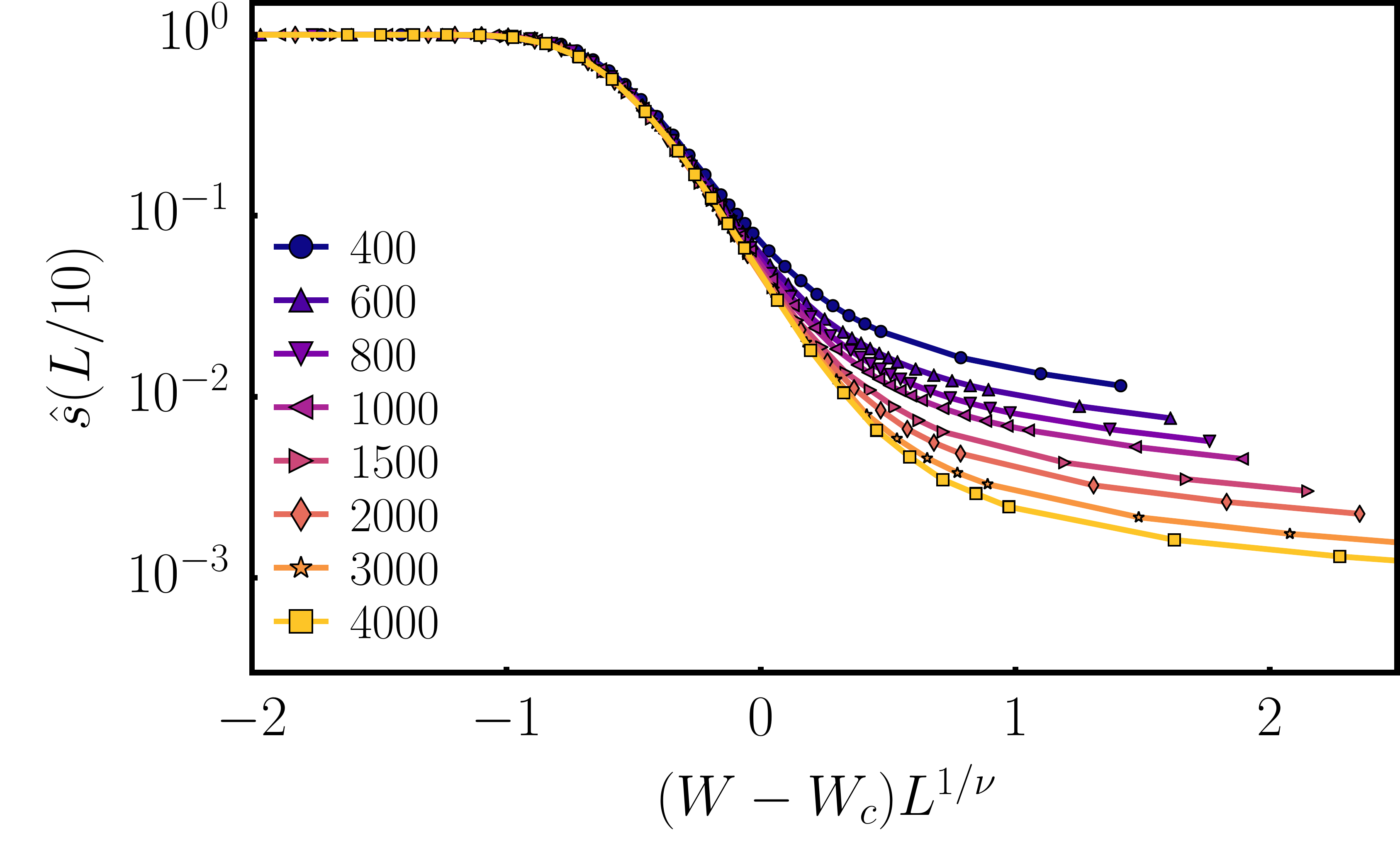}
\caption{
Scaling collapse of entanglement $\hat{s}(Lf)$ for different system fractions $f = x/L$. Here $W_c = 2.05$ and $\nu = 3.2$, indicating a universal form of the volume law coefficient \eqref{eq:univVolLaw}. The $f = 1/2$ scaling collapse is shown in Fig.~\ref{fig:EntanglementCrossingL2} (inset). As expected, the $L$ curves start separating due to finite size corrections sooner for smaller $f$. Nonetheless, even for $f = 1/10$, the collapse of $L = 3000, 4000$ persists somewhat for $W > W_c$. The $L=4000$ data are those shown in Fig.~\ref{fig:S-x-dependence}(b).
}
\end{figure*}

\newpage
\onecolumngrid

\begin{figure}[ht]
\raggedright
\includegraphics[width=0.32\columnwidth]{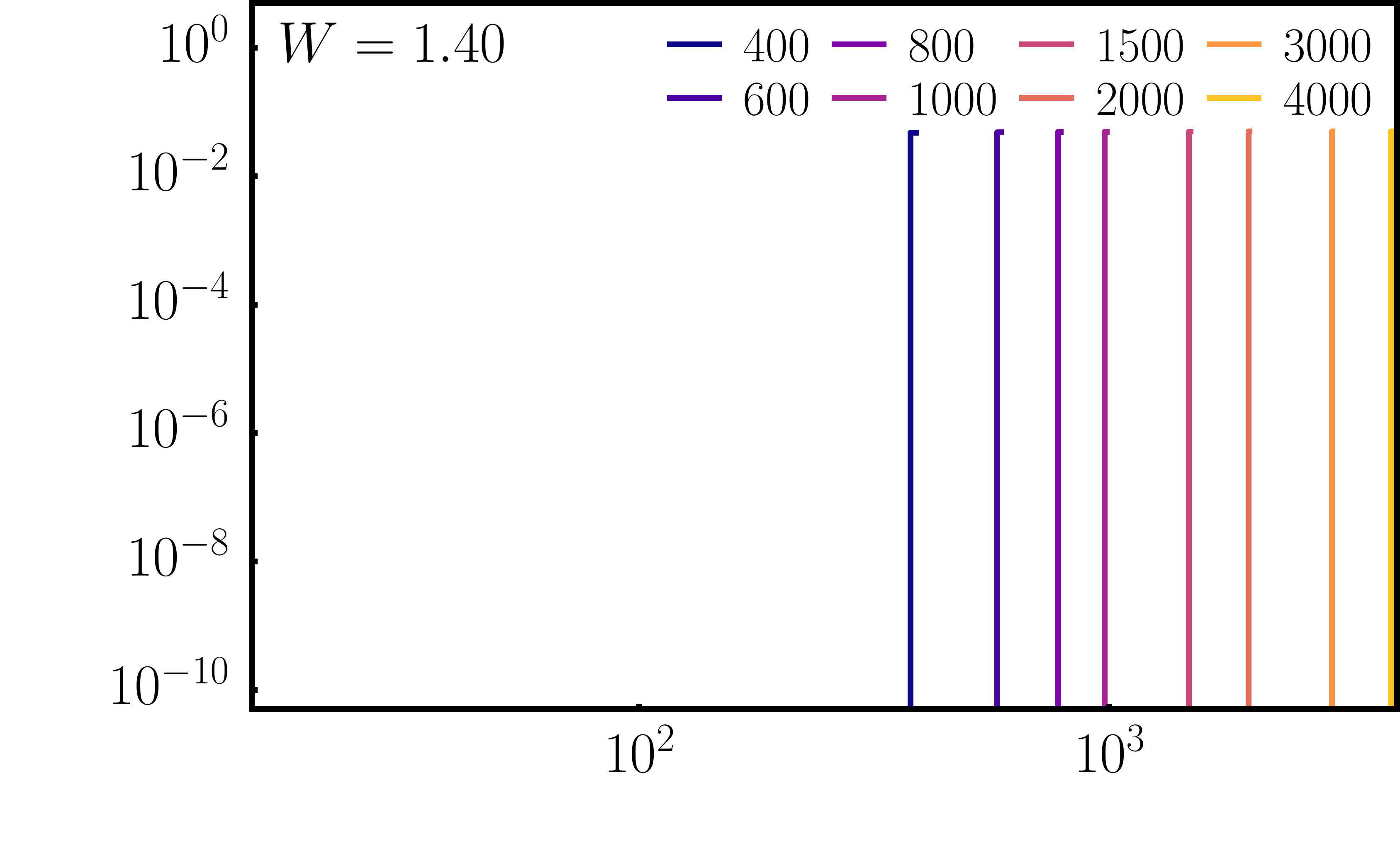}
\includegraphics[width=0.32\columnwidth]{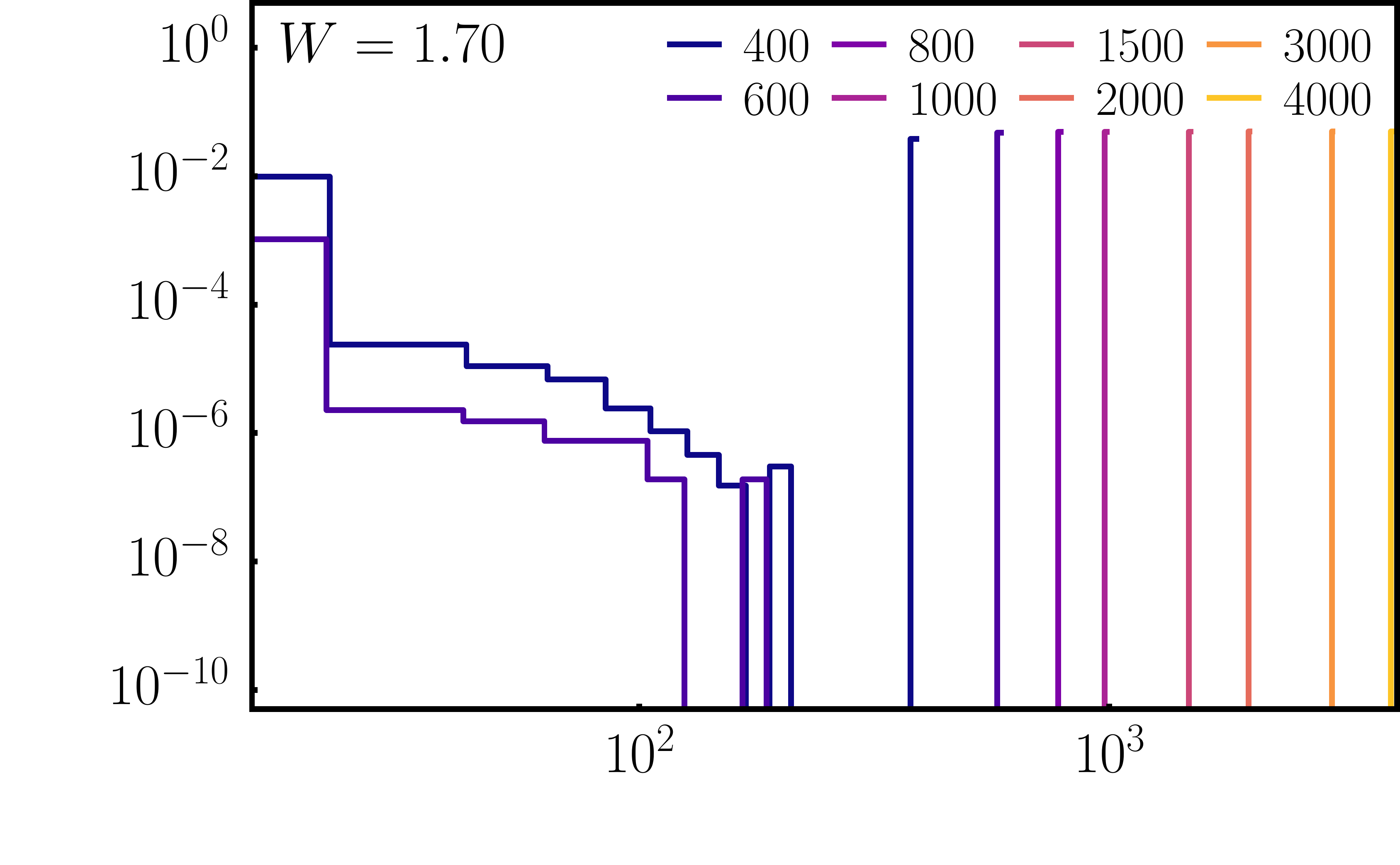}
\includegraphics[width=0.32\columnwidth]{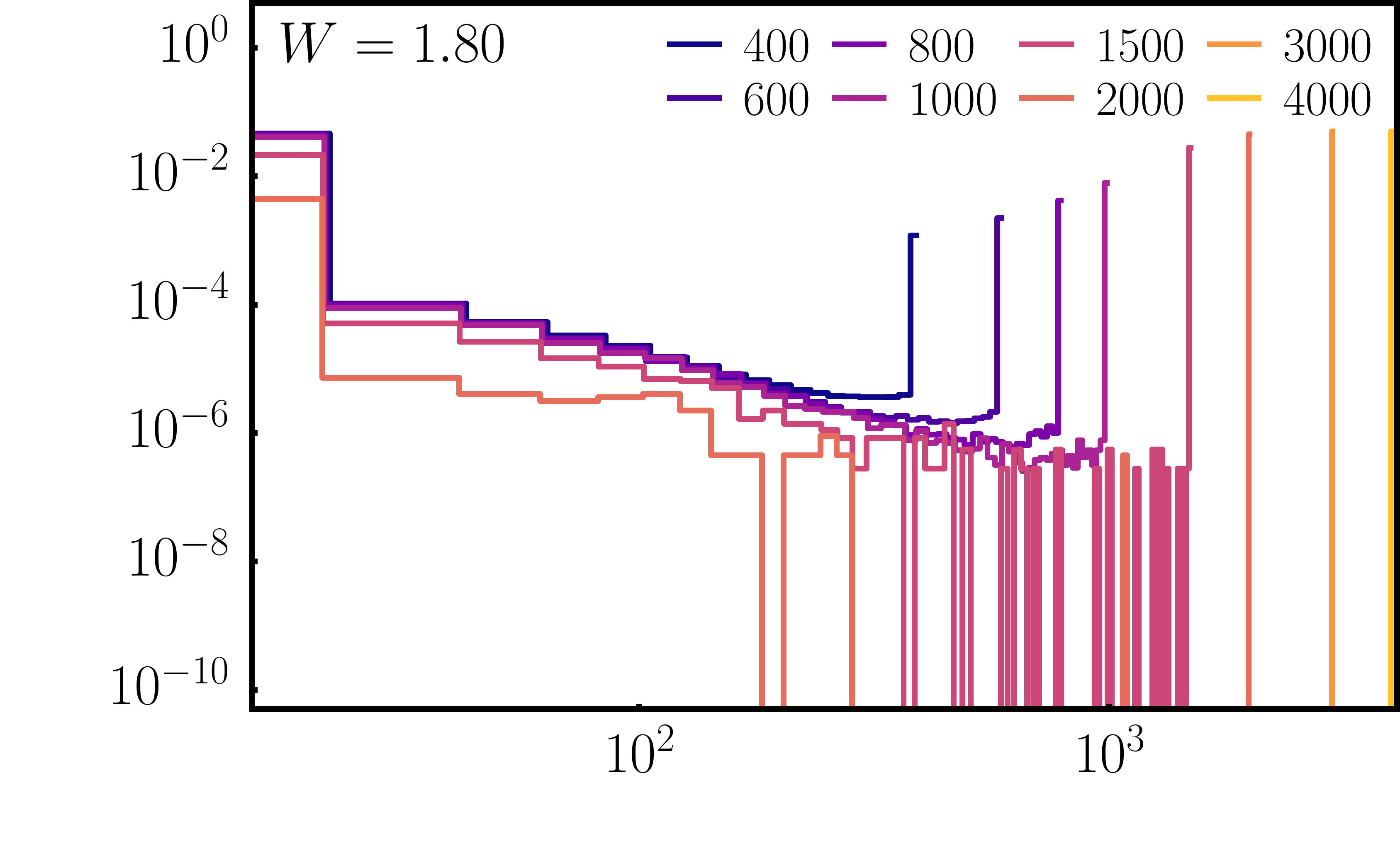}
\includegraphics[width=0.32\columnwidth]{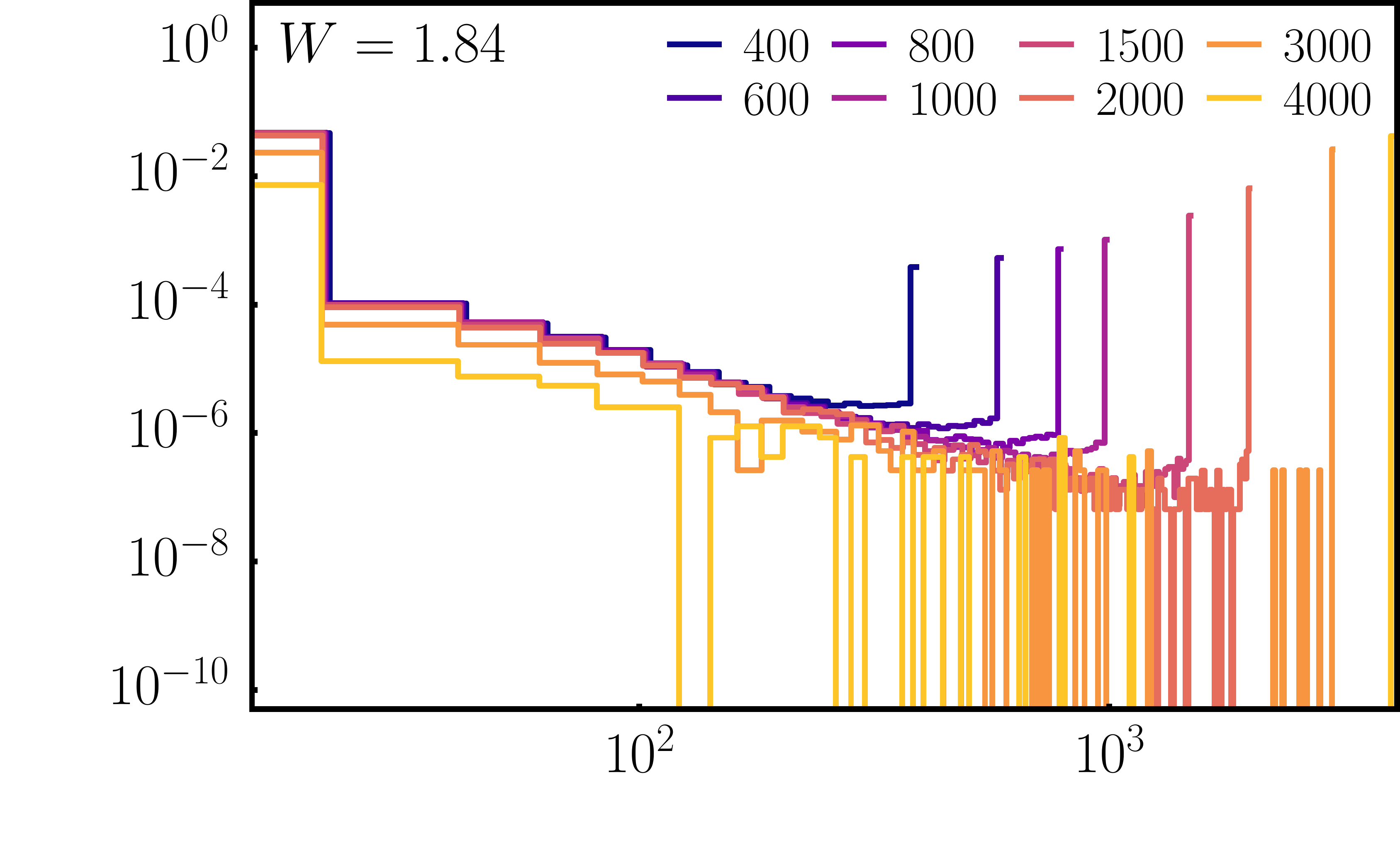}
\includegraphics[width=0.32\columnwidth]{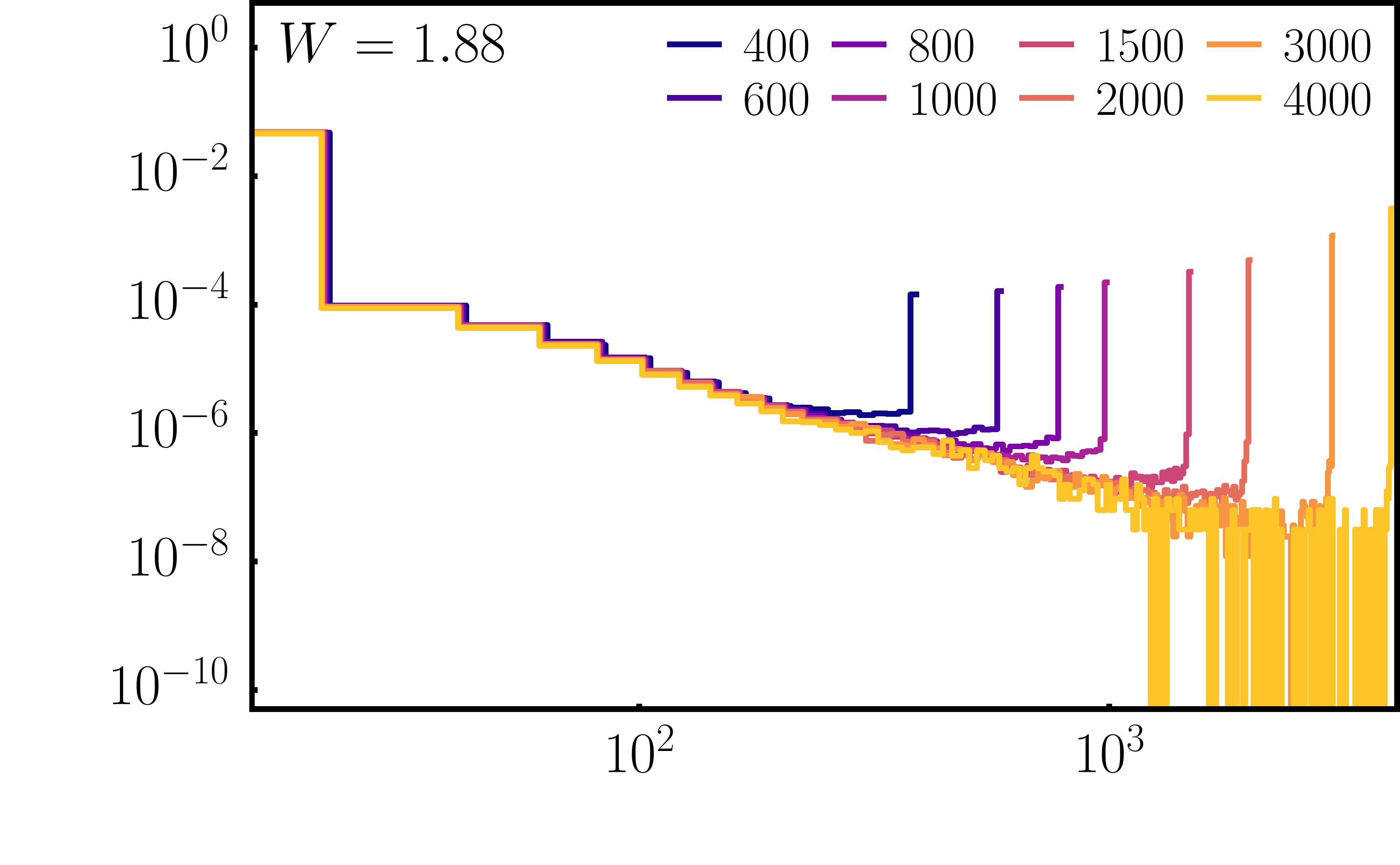}
\includegraphics[width=0.32\columnwidth]{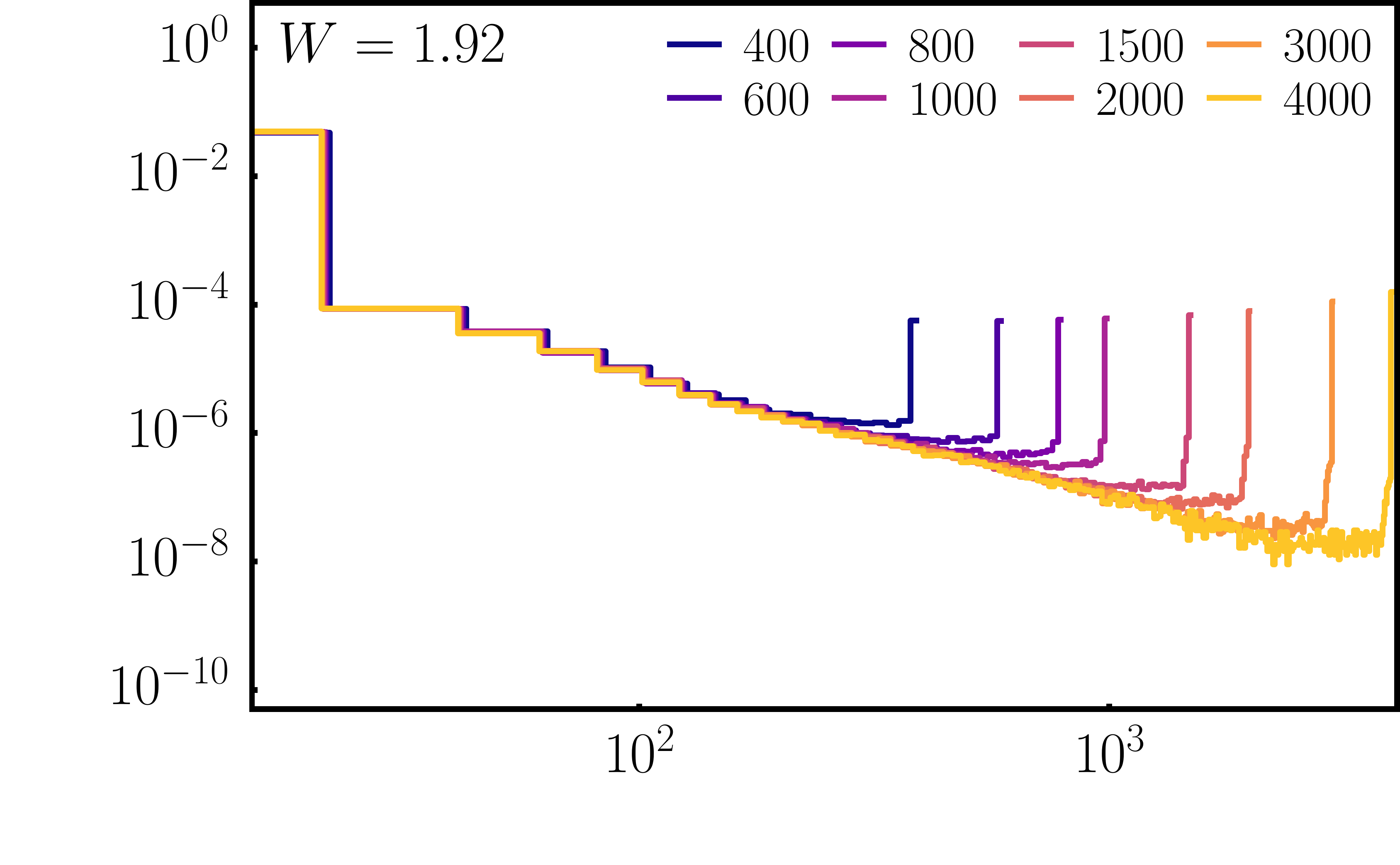}
\includegraphics[width=0.32\columnwidth]{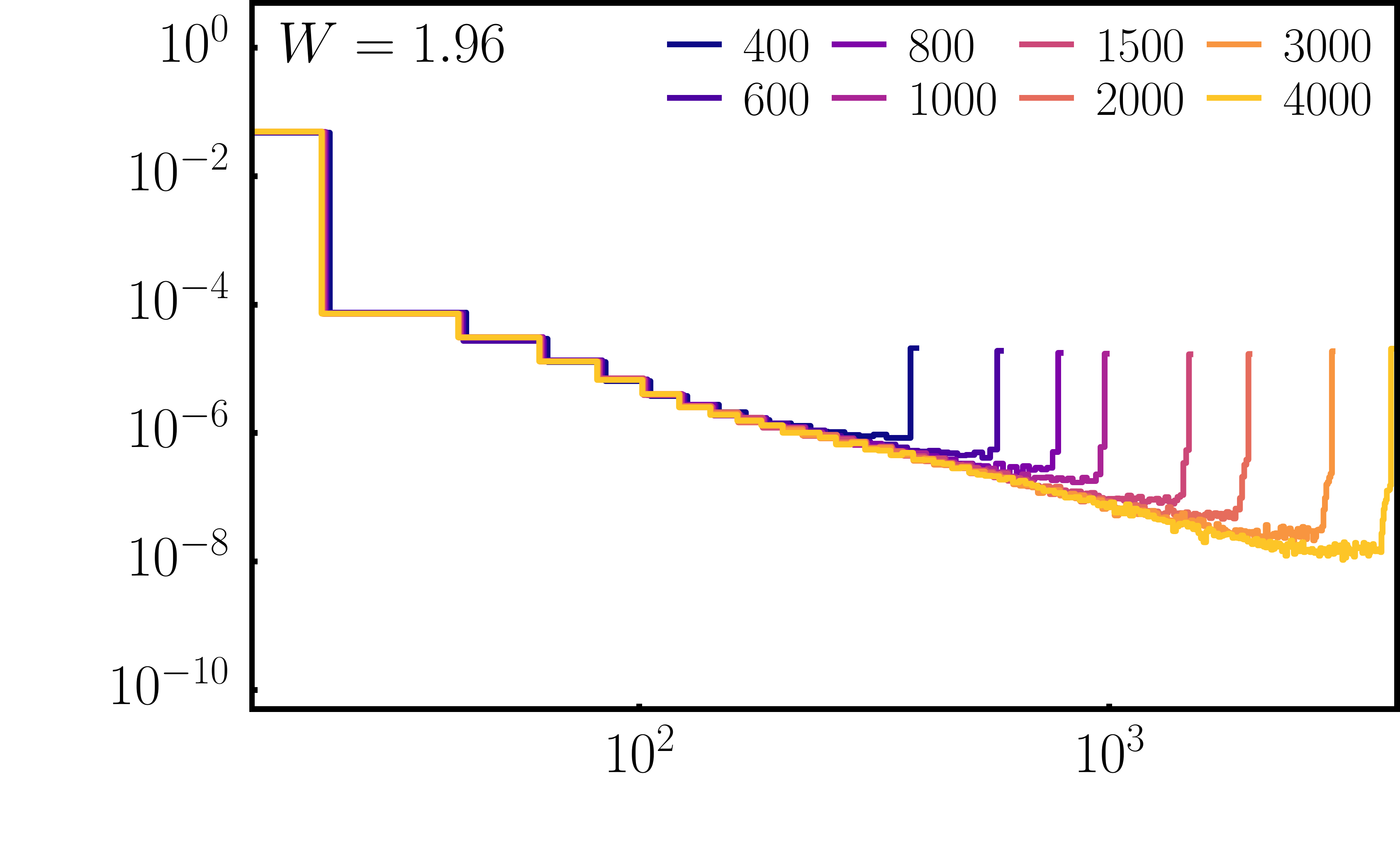}
\includegraphics[width=0.32\columnwidth]{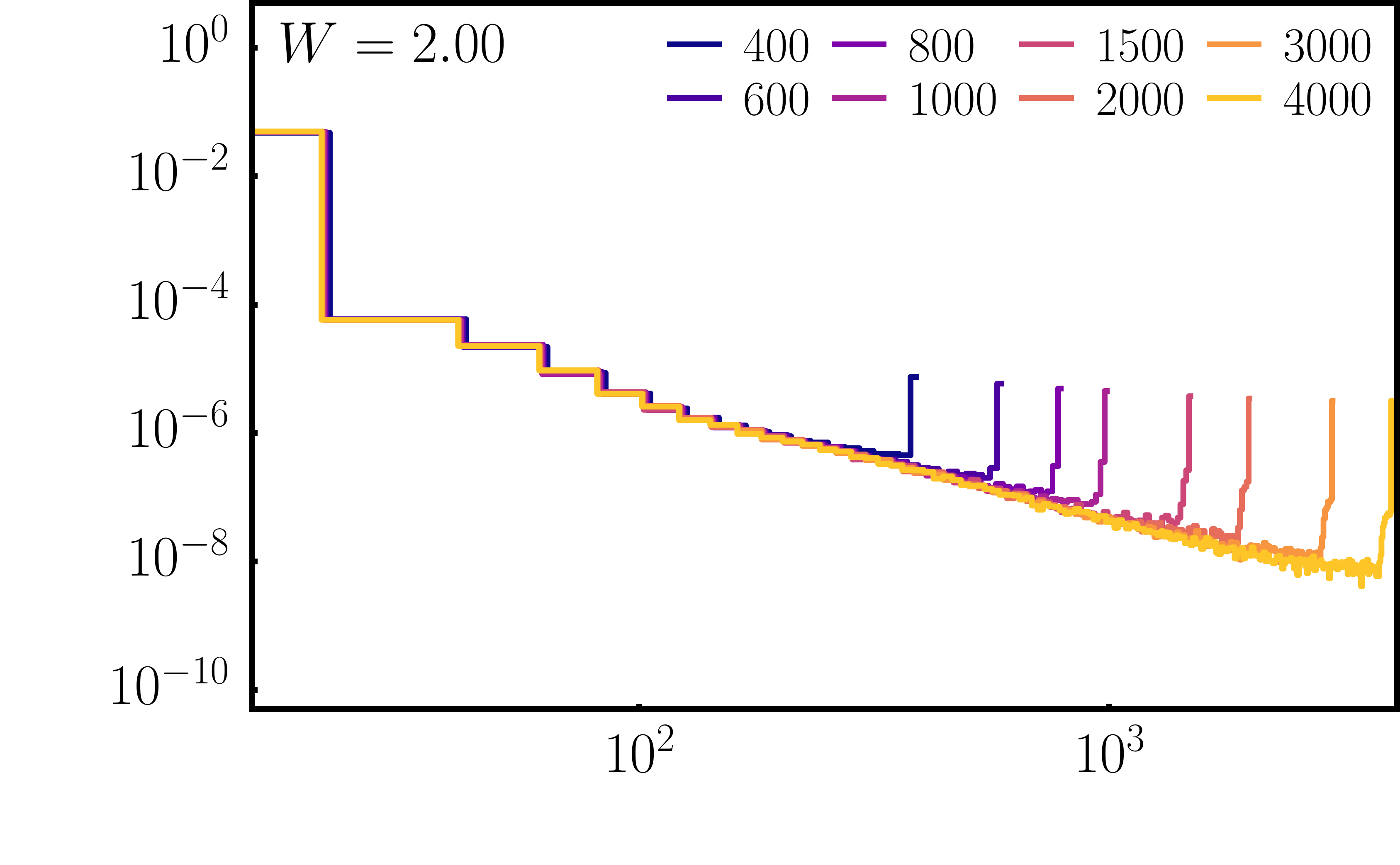}
\includegraphics[width=0.32\columnwidth]{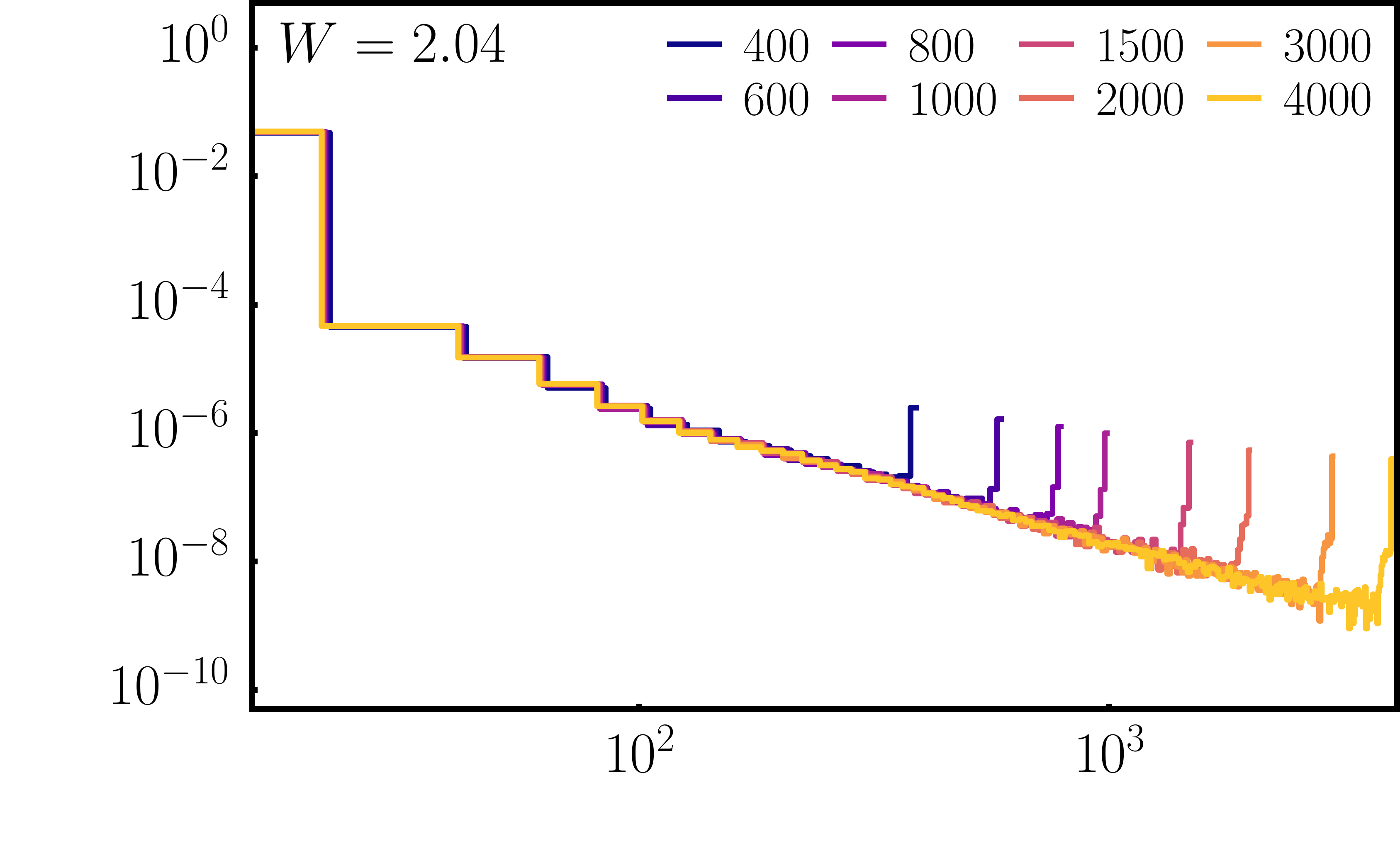}
\includegraphics[width=0.32\columnwidth]{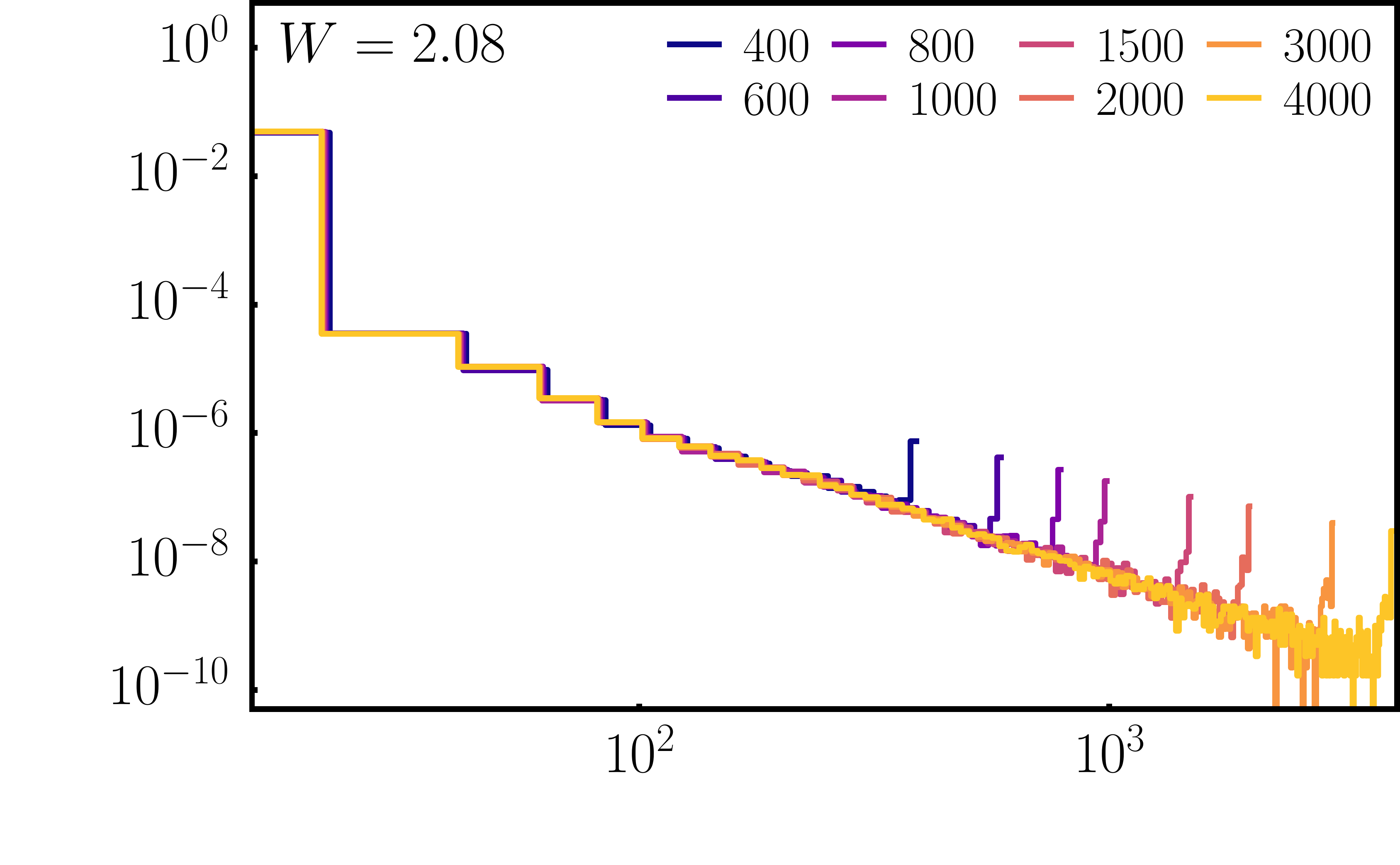}
\includegraphics[width=0.32\columnwidth]{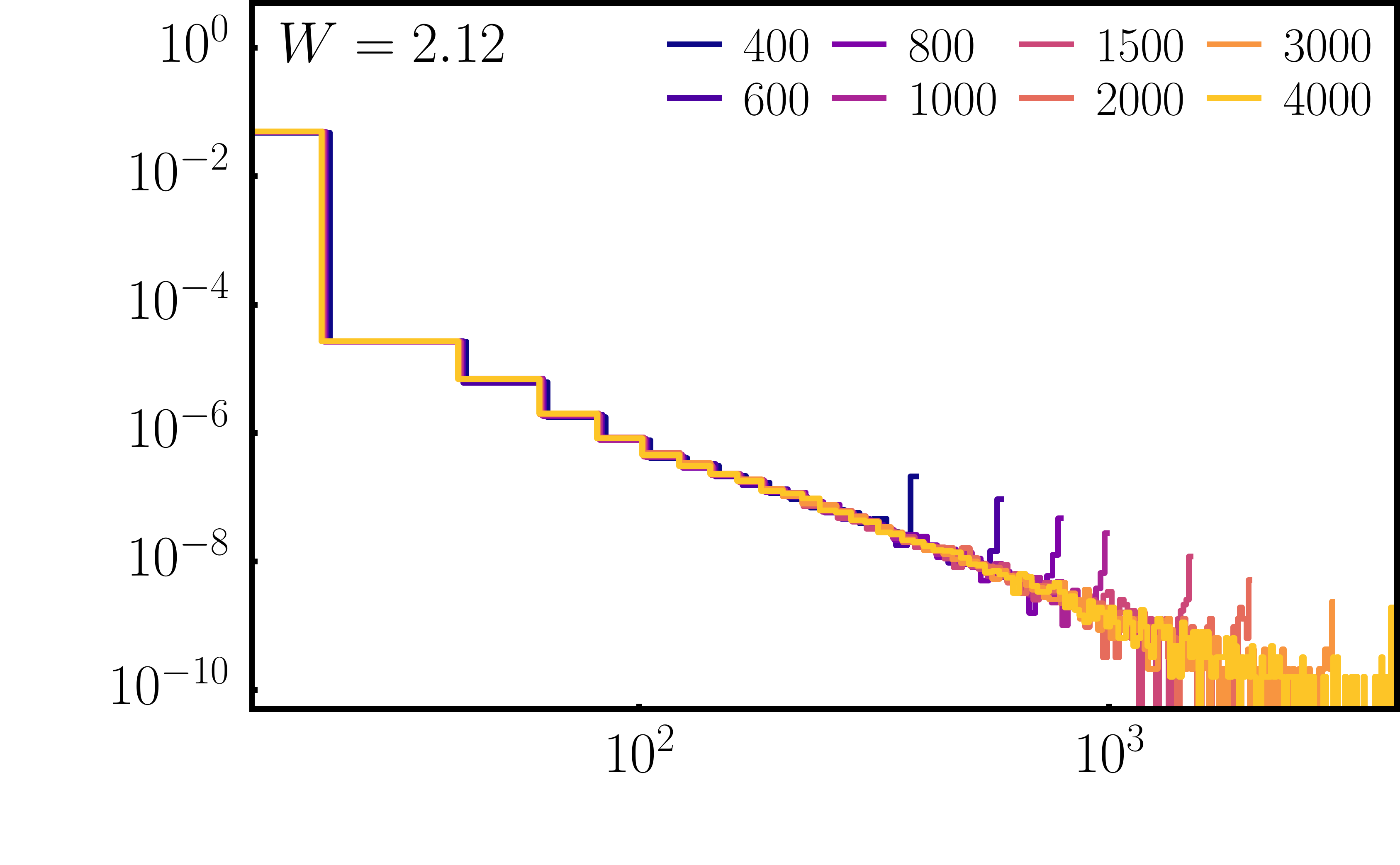}
\includegraphics[width=0.32\columnwidth]{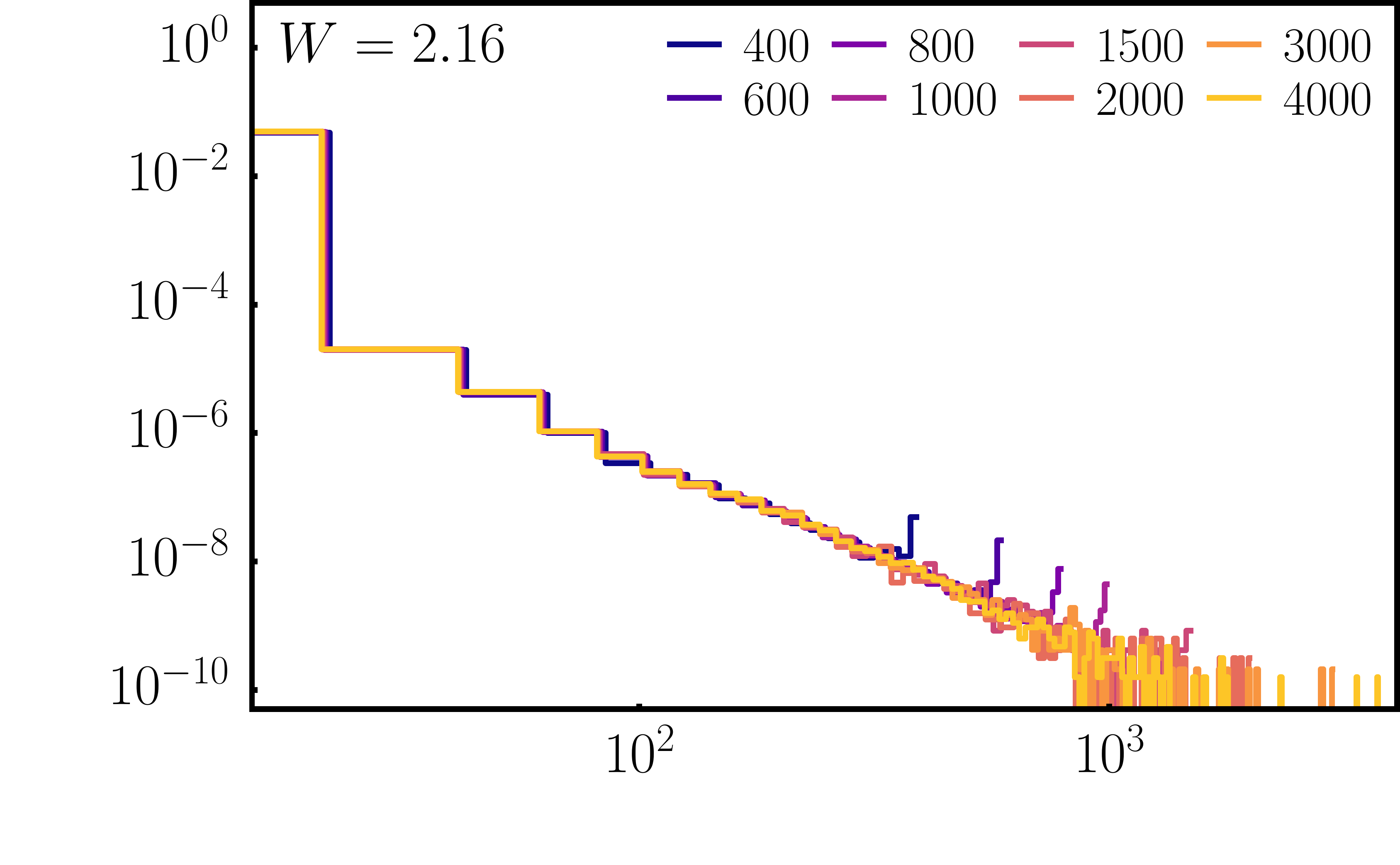}
\includegraphics[width=0.32\columnwidth]{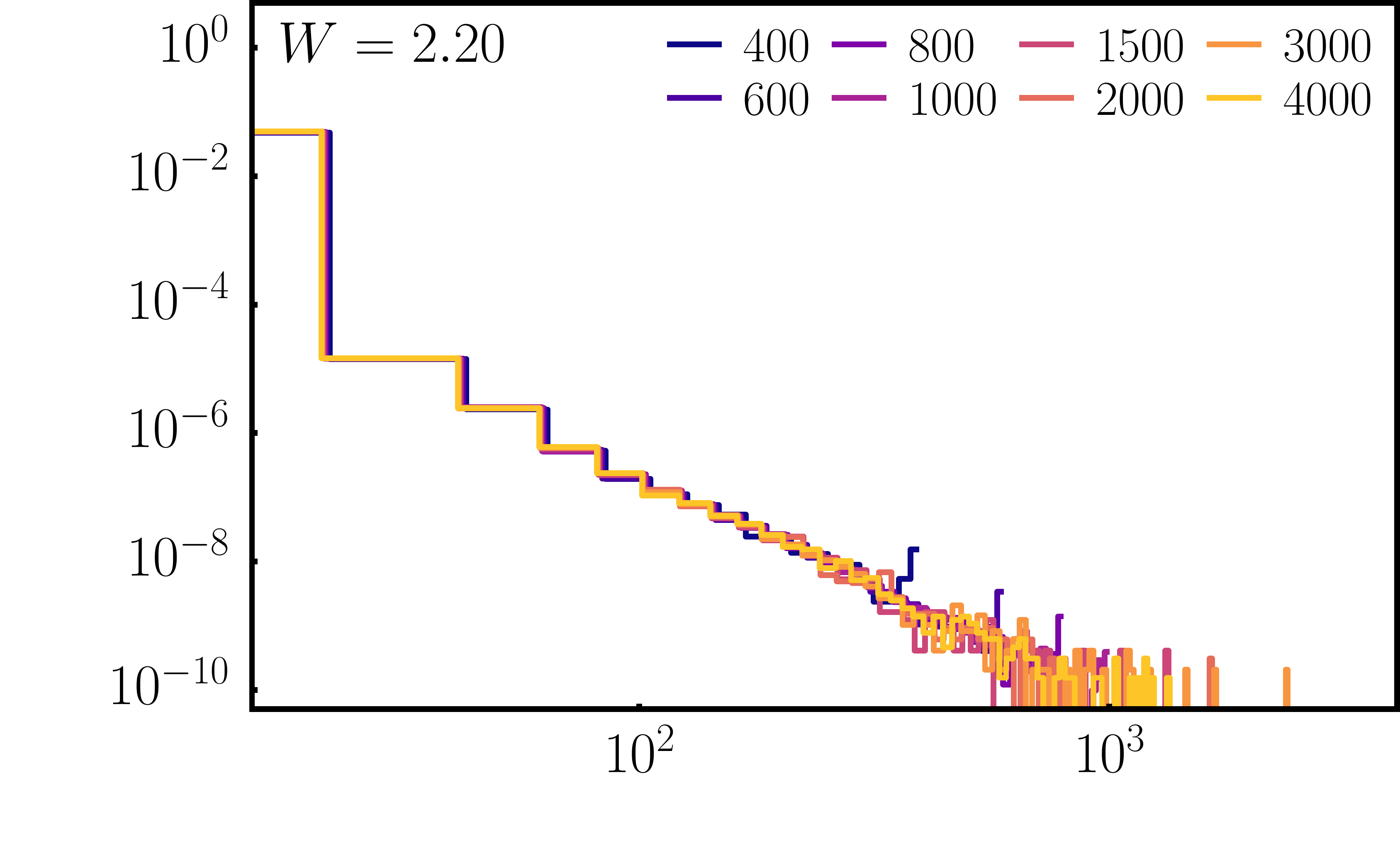}
\includegraphics[width=0.32\columnwidth]{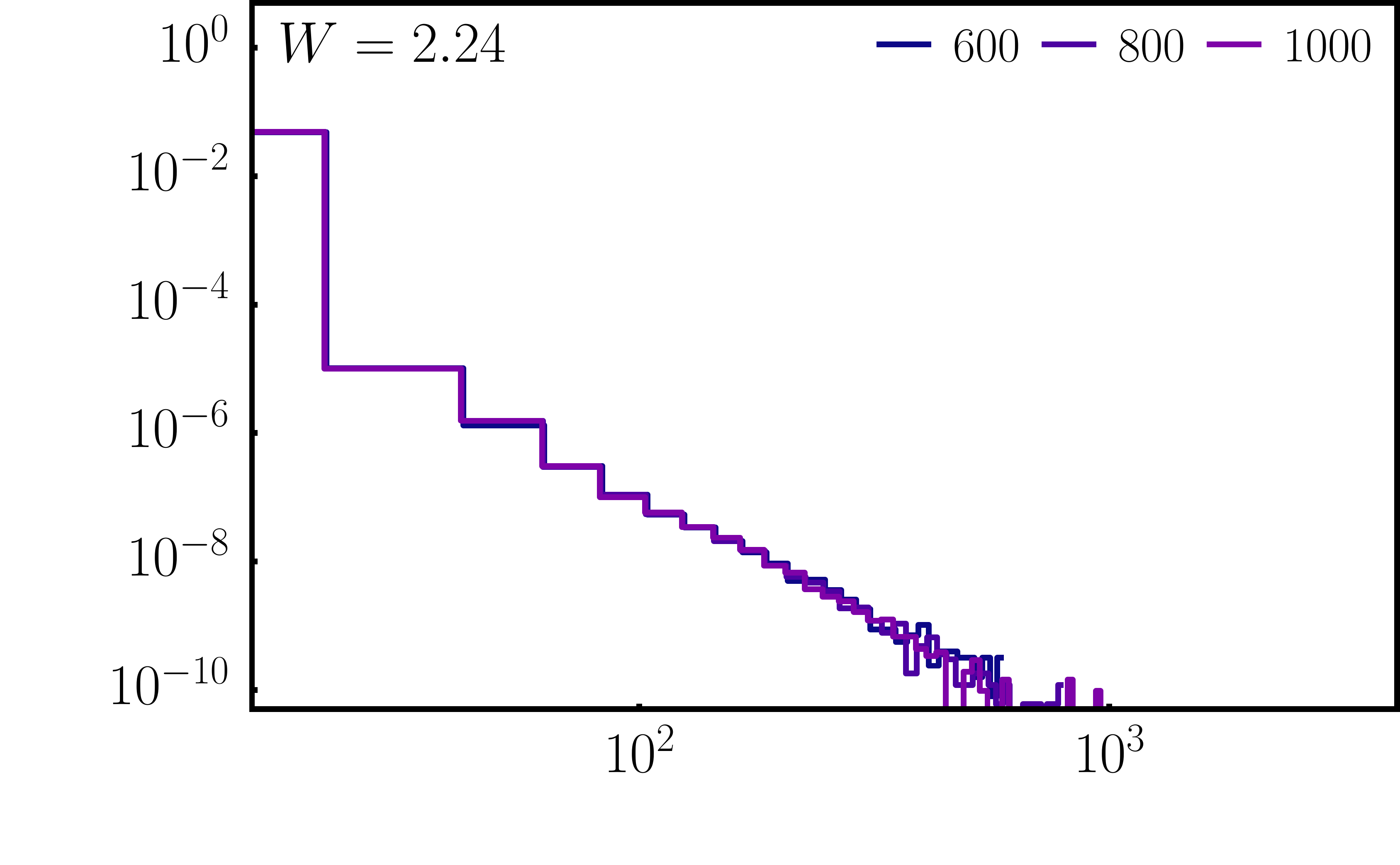}
\includegraphics[width=0.32\columnwidth]{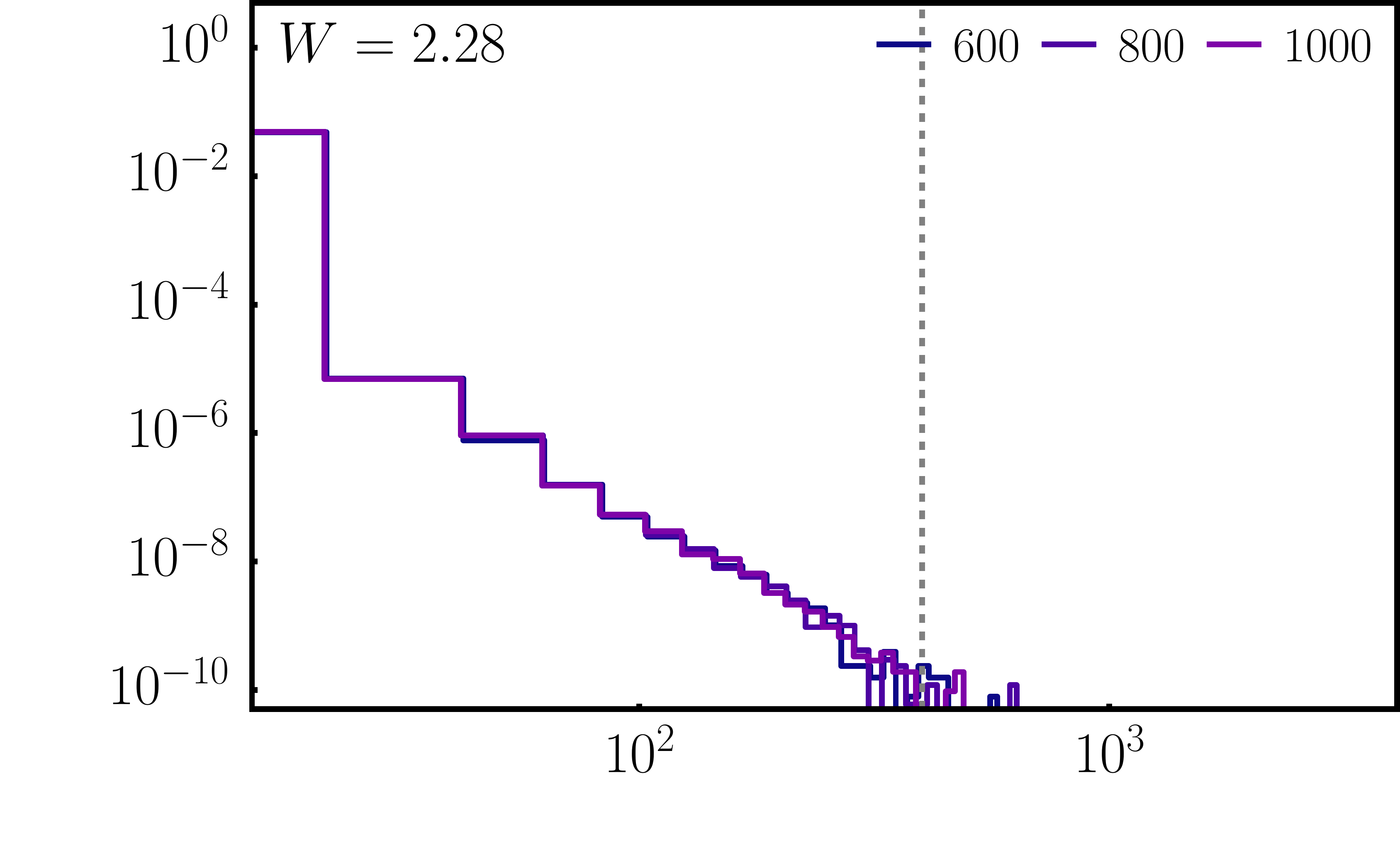}
\includegraphics[width=0.32\columnwidth]{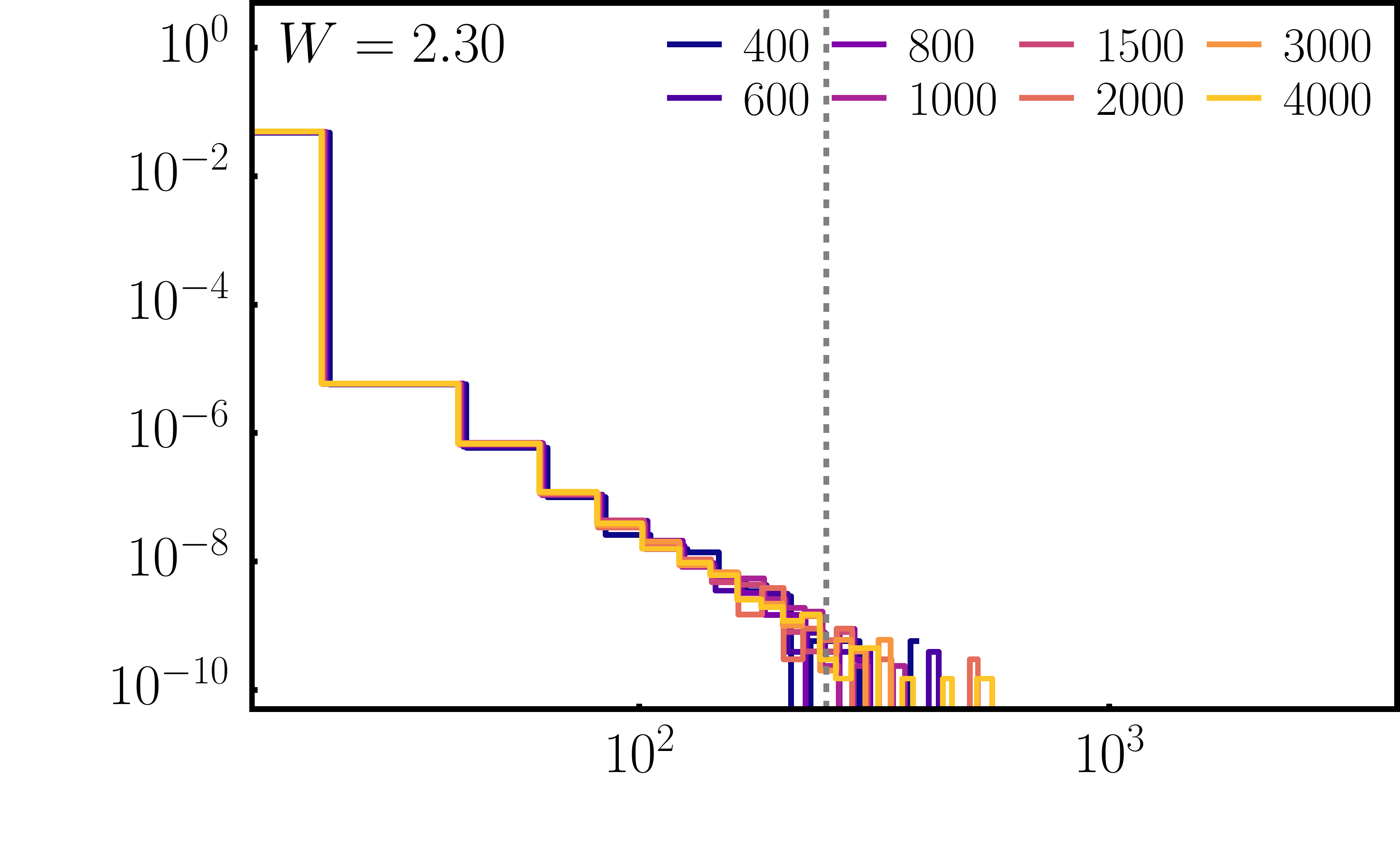}
\includegraphics[width=0.32\columnwidth]{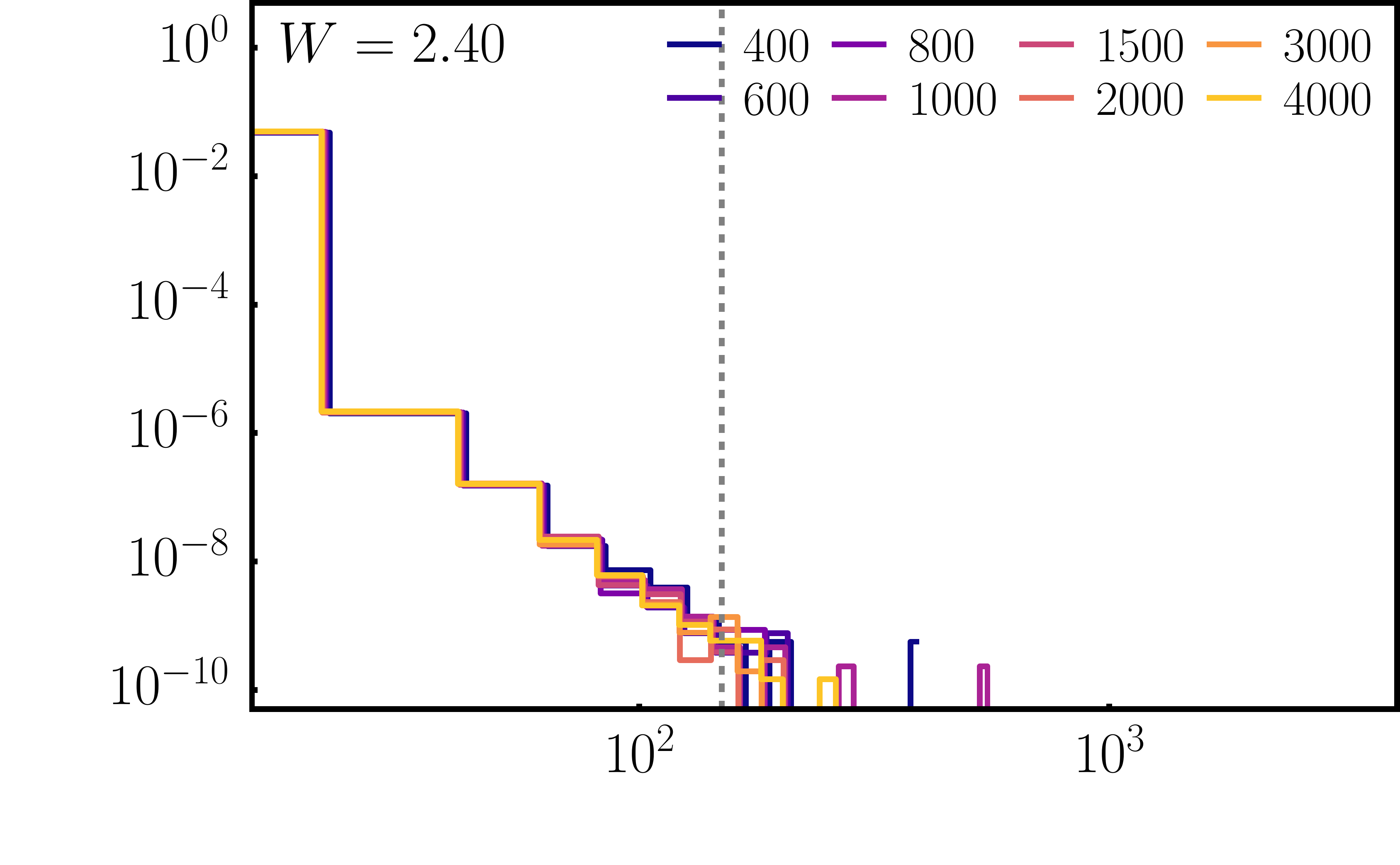}
\includegraphics[width=0.32\columnwidth]{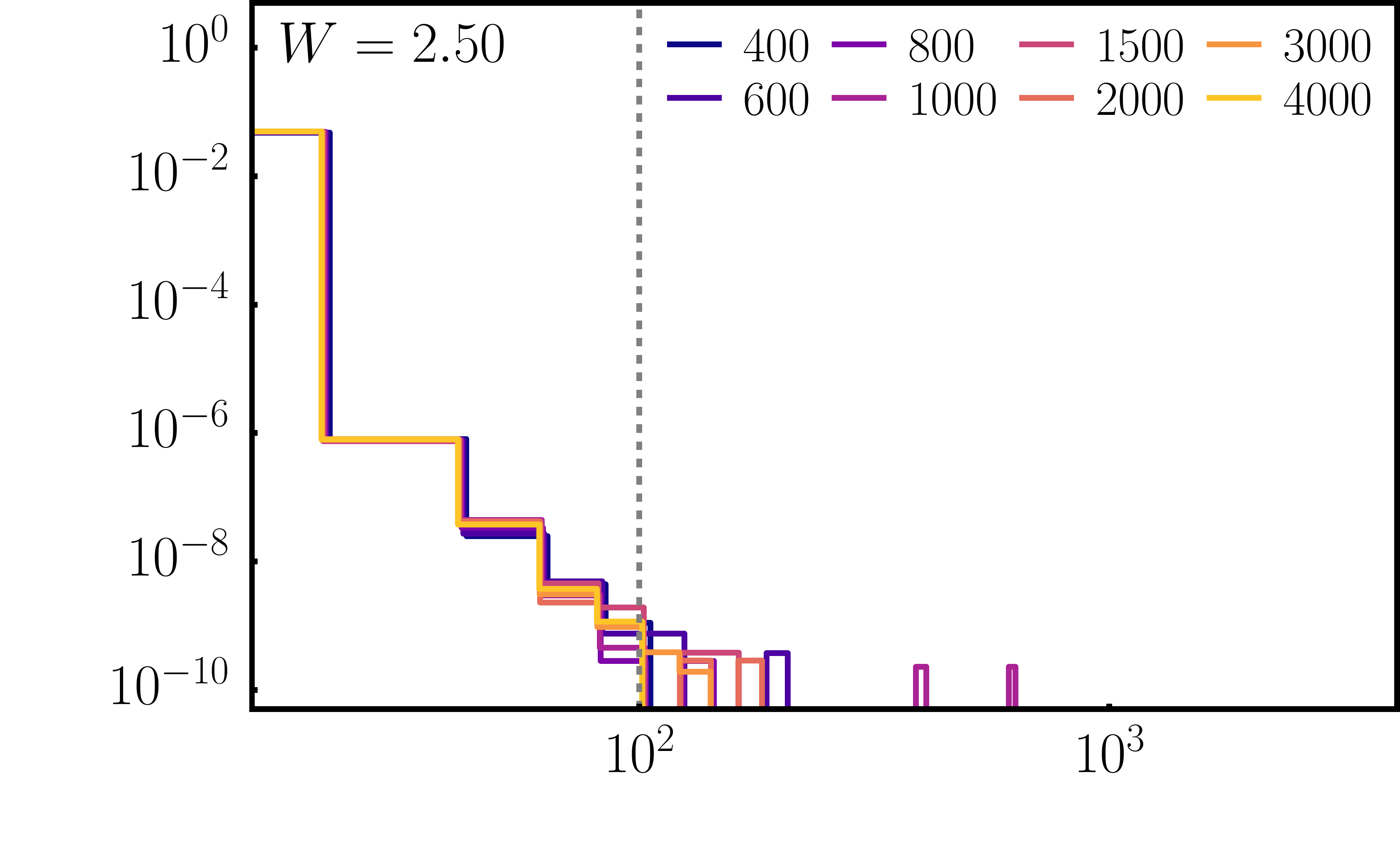}

\caption{
Normalized histogram over all cluster sizes realized at a given disorder bandwidth $W$ for different system sizes $L$.  Deep in the thermal phase (e.g.~$W=1.4$), each disorder realization consists of one thermal cluster spanning the system.  Near the critical point the distribution is bimodal -- there is a finite thermal peak and a universal power law. As we move away from the critical point the power law changes to an exponential decay at a length-scale which we can identify as the correlation length $\xi$. As we care to establish the asymptotic regime $1 \gg x \gg \xi \gg L$, extracting lower bounds on $\xi$ is sufficient. These are given as vertical dashed lines for those $W$ marked in Fig.~\ref{fig:S-x-dependence}(a).  The histograms use $L/20$ bins.}
\end{figure}

\end{document}